\documentclass{pasj01}

\usepackage{color}
\usepackage{natbib}

\Received{$\langle$reception date$\rangle$}
\Accepted{$\langle$acception date$\rangle$}
\Published{$\langle$publication date$\rangle$}
\begin{document}

\title{A survey of molecular cores in M17 SWex}

%%% begin:list of authors
\author{Tomomi \textsc{Shimoikura}\altaffilmark{1,2}, Kazuhito \textsc{Dobashi}\altaffilmark{1},
Asha \textsc{Hirose}\altaffilmark{1},
Fumitaka \textsc{Nakamura}\altaffilmark{3,4}, Yoshito \textsc{Shimajiri}\altaffilmark{3,5,6}, and Koji \textsc{Sugitani}\altaffilmark{7}}
\altaffiltext{1}{Department of Astronomy and Earth Sciences, Tokyo Gakugei University, Koganei, Tokyo  184-8501, Japan} 
\altaffiltext{2}{Faculty of Social Information Studies, Otsuma Women's University, Chiyoda-ku,Tokyo, 102-8357, Japan} 
\altaffiltext{3}{National Astronomical Observatory of Japan, Mitaka, Tokyo 181-8588, Japan}
\altaffiltext{4}{Department of Astronomical Science, School of Physical Science, SOKENDAI (The Graduate University for Advanced Studies), Osawa, Mitaka, Tokyo 181-8588, Japan}
\altaffiltext{5}{Laboratoire AIM, CEA/DSM-CNRS-Universit$\acute{\rm e}$ Paris Diderot, IRFU/Service d'Astrophysique, CEA Saclay, F-91191 Gif-sur-Yvette, France}
\altaffiltext{6}{Department of Physics and Astronomy, Graduate School of Science and Engineering, Kagoshima University, 1-21-35 Korimoto, Kagoshima, Kagoshima, 890-0065, Japan}
\altaffiltext{7}{Graduate School of Natural Sciences, Nagoya City University, Mizuho-ku, Nagoya 467-8501, Japan}

\email{ikura@otsuma.ac.jp}
%%% end:list of authors


\maketitle

%%%%%%%%%%%%%%%%%%%%%%%%%%%%%%%%%%%
%                                                      Abstract                                 %
%%%%%%%%%%%%%%%%%%%%%%%%%%%%%%%%%%%

\begin{abstract}
A survey of molecular cores covering the infrared dark cloud known as the M17 southwest extension (M17 SWex) has been carried out with the 45 m Nobeyama Radio Telescope.
Based on the N$_2$H$^{+}$ ($J=1-0$) data obtained, 
we have identified 46 individual cores whose masses are in the range 43 to 3026 ${M}_{\odot}$.
We examined the relationship between the physical parameters of the cores and those of young stellar objects (YSOs) associated with the cores found in the literature.
%{\color{blue}{We estimated the average star formation efficiency of M17 SWex to be $\sim9-17\%$, which is comparable to those of other cluster-forming clumps producing massive stars.}}
%There is a correlation between the core mass estimated from the Herschel data and the total stellar mass associated with each core.
The comparison of the virial mass and the core mass indicates 
that most of the cores can be gravitationally stable if we assume a large external pressure. 
%The results of the comparison also suggest that internal pressure due to turbulence for some cores is small and they are contracting.
%{\color{red}{In addition, investigating the density structures of the cores suggests that many cores are infalling.}}
Among the 46 cores, we found four massive cores with YSOs. 
They have large mass of $\gtsim1000M_{\odot}$ and line width of $\gtsim 2.5$ km s$^{-1}$
which are similar to those of clumps forming high mass stars. 
%but they are not associated with apparent high-mass stars.
However, previous studies have shown that there is no active massive star formation in this region.
Recent measurements of near-infrared  polarization infer that the magnetic field around M17 SWex
is likely to be strong enough to support the cores against self-gravity.
We therefore suggest that the magnetic field may prevent the cores from collapsing,
causing the low-level of massive star formation in M17 SWex.
%We concluded that the cores in M17 SWex may not form high-mass stars 
%because of the magnetic field reported in the previous studies.
\end{abstract}

%%%%%%%%%%%%%%%%%%%%%%%%%%%%%%%%%%%
%                                         INTRODUCTION                                  %
%%%%%%%%%%%%%%%%%%%%%%%%%%%%%%%%%%%

\section{Introduction }
The infrared dark cloud M17 southwest extension \citep[M17 SWex,][]{Povich2010}, which is shaped like a ``flying dragon" is located to the southwest of the M17 H{$\,${\sc ii}} region.
A huge ($\sim10^{6} M_\odot$) molecular cloud was found by the CO observations in the region using the Texas 5 m telescope with a beam size of $\sim3\arcmin$ \citep{Elmegreen}. 
More recently, \cite{Busquet} conducted NH$_3$ observations using the Very Large Array and Effelsberg 100 m telescope 
toward a part of the high density region in M17 SWex,
and they revealed a network of filaments constituting two hub-filament systems.
The two hubs are dubbed ``hub-N" and ``hub-S".
\cite{Busquet} suggested that they are the main sites of stellar activity within the cloud.

\cite{Povich2016} suggested that M17 SWex has been an active star-forming region for the past $\gtsim1$ Myr.
\cite{Povich2010} identified 488 Young Stellar Objects (YSOs) in M17 SWex, 
and about 200 of the YSOs are found to be stars with a mass greater than $3M_\odot$ that will grow into B-type stars in the future.
They suggested that M17 SWex probably has not yet formed its most massive star, predicted to be an early O-type star.
From the 3 mm continuum emission observations toward the two hubs with ALMA, \cite{Ohashi2016} identified 48 cores and indicated that the internal
turbulence is insufficient to prevent the gravitational collapse of the two hubs.
They also suggested that clumps hosting the cores may be able to supply material to the cores.
\cite{Povich2016}  and  \cite{Ohashi2016} speculate that mass accretion onto the cores via the hubs
could produce massive star formation in the future. 
M17 SWex is thus considered to be in the early stages of formation of a new OB association \citep[e.g.,][]{Elmegreen, Povich2010}.
Therefore, M17 SWex is a molecular cloud suitable for study to explore the initial state of massive star formation.
However, high-resolution observations of molecular emission lines in this region have been done 
only toward a limited area including hub-N and hub-S. 
The Millimeter Astronomy Legacy Team 90 GHz (MALT90) survey \citep{Foster2011} revealed some dense cores in M17.
A comprehensive molecular core survey in the cloud to cover the entire M17 SWex region is needed.

In other star forming regions, \cite{Shimoikura2018} found massive
($\sim10^{3}M_{\odot}$) and dense ($\sim10^{5}$ cm$^{-3}$ ) clumps with no massive stars.
They suggested that such clumps are gravitationally stable without collapsing due to clump-supporting 
forces such as turbulence and magnetic fields.
\cite{Sugitani2019} conducted near-infrared polarization observations toward M17 SWex. 
They revealed filament-like structures in the H$_2$ column density map, 
and found that the local magnetic field is perpendicular to 
most of the individual filamentary structures in high column density regions.
They suggested that the magnetic field is likely to influence the formation and evolution of M17 SWex.
To understand the relationship between the cores and the magnetic field in M17 SWex,
it is necessary to identify dense cores in the region and to investigate the dynamical stability of the cores.

To search for dense cores in M17 SWex and to study their dynamics, 
we observed the region using the 45 m radio telescope at Nobeyama Radio Observatory (NRO) 
in some molecular lines at 93-115 GHz. 
In this study, we aim to identify the molecular cores in M17 SWex and catalog them using the N$_2$H$^{+}$ molecular emission line at 93 GHz, which is a good tracer of dense gas.
We investigate physical parameters of the cores and 
examine the relationship between physical properties of the cores and those of YSOs associated with the cores.

\cite{Povich2010} assumed that M17 SWex 
is at a distance of 2.0 kpc measured in the M17 H{$\,${\sc ii}} region based on the parallax measurements by \cite{Xu2011},
since the LSR velocity $V_{\rm{LSR}}$ of M17 SWex and the H{$\,${\sc ii}} region are the same
($V_{\rm{LSR}}\simeq 20$ km s$^{-1}$). 
In our study, we also found that
M17 SWex is likely to be connected to the H{$\,${\sc ii}} region,
and thus we adopt 2.0 kpc as the distance to M17 SWex
in this paper.

This study is based on ``the Star Formation Legacy project" 
which is a large-scale survey of molecular gas in star forming regions. 
The outline of the project is presented by Nakamura et al. (2019a).
Results of other regions are given in separate articles 
(OrionA: Nakamura et al. 2019b, Ishii et al. 2019, Tanabe et al. 2019,
% H. Takemura et al. in preparation, 
Aquila Rift: Shimoikura et al. 2019, Kusune et al. 2019, M17: Sugitani et al. 2019, 
NCS: Dobashi et al. 2019a, DR21: Dobashi et al. 2019b).

%\cite{Busquet} showed that the two hubs show larger velocity dispersion and larger masses per unit length than filaments. 
%So far, various observations have been made by other researchers toward the only these two hubs.
%N2H+ traces the high density regions of dark clouds as it is less depleted onto dust grain surfaces than such as the CO molecule. Thus, it can be found in prestellar and protostellar clumps (Bergin and Langer, 1997;Hotzel et al., 2004; Daniel et al., 2006).

%%%%%%%%%%%%%%%%%%%%%%%%%%%%%%%%%%%
%                                        OBSERVATIONS                                      %
%%%%%%%%%%%%%%%%%%%%%%%%%%%%%%%%%%%

\section{Observations }
 
\subsection{Observations with the NRO 45 m telescope }

Observations of the $^{12}$CO, $^{13}$CO, C$^{18}$O, CCS, and N$_{2}$H$^{+}$ emission lines were carried out 
with the 45 m telescope at NRO.
We observed the $^{12}$CO and $^{13}$CO emission lines for 20 hours in the period between 2015 April and 2016 March, 
and the other emission lines for 50 hours in the period between 2016 April and 2017 March. 
We used an on-the-fly (OTF) observing technique that was implemented for the 45 m telescope by \cite{Sawada}.
We observed whole of the M17 H{$\,${\sc ii}} region and M17 SWex ($\sim1.5^{\circ}\times0.5^{\circ}$) 
in the $^{12}$CO and $^{13}$CO lines.
We also mapped an $\sim1^{\circ}\times0.5^{\circ}$ area in M17 SWex in the other lines.
The beam size of the 45 m telescope is $\sim15\arcsec$ (HPBW) at 100 GHz.
We used the multi-beam receiver “FOREST” \cite[FOur beam REceiver System on the 45 m Telescope;][]{Minamidani} as the frontend and the digital spectrometer SAM45 as the backend.
The velocity resolution was set to $0.04$ km s$^{-1}$ for the observed emission lines. 
The receiver provided a typical system temperature of 170 K.
%The spectrometer was digital spectrometer with 4096 channels, and the total bandwidth and the channel separation were 31.25 MHz and 15.26 kHz, respectively. 
%Calibration was done by the chopper-wheel technique (Kutner & Ulich 1981) to get line strengths on %the antenna temperature ($T^*_{\rm A}$) scale.
The pointing was checked every two hours by observing the SiO maser source V1111-Oph and was accurate within $5\arcsec$.
The intensity calibration was made by observing a small region of $\sim1\arcmin\times1\arcmin$ in hub-N every time we tuned the receiver, 
and we found that the intensity fluctuations for all of the lines are less than $10\%$.

The data reduction for baseline subtraction was carried out using the NOSTAR software package developed at NRO.
%A map was generated by convolving the observed data with a spheroidal function and regridding with an 7.5 %grid along the spatial direction and a velocity resolution of 0.1kms.
The 3D fits data were generated by convolving the observed data with a spheroidal function 
and regridding them at $7\farcs5$ per pixel, resulting in an effective angular resolution of $22\arcsec - 24\arcsec$, 
corresponding to a linear resolution 0.2 pc at a distance of 2.0 kpc.
To increase the signal-to-noise ratios, we regridded the data to a velocity resolution of 0.1 km s$^{-1}$.
We converted the antenna temperature $T_{\rm a}^{*}$ to the main beam temperature $T_{\rm{mb}}$ assuming that the main beam efficiency of the telescope is 0.416, 0.435, 0.437, 0.497, and 0.500 
for the $^{12}$CO, $^{13}$CO, C$^{18}$O, CCS, and N$_2$H$^{+}$ emission lines, respectively.
The rms noise of the final data is $\Delta T_{\rm{mb}}=0.4 - 1.0$ K.
We summarize
the observed molecular lines and the resulting noise levels in table \ref{tab:line}.
More detailed description for the data reduction are summarized by \cite{Nakamura2019a} and \cite{Shimoikura2019}.

\subsection{Archival data}
We used the Herschel archival data of 160, 250, 350, and 500 $\micron$ toward M17 SWex 
to construct an H$_2$ column density $N$(H$_2$) map and a dust temperature $T_{\rm{dust}}$ map 
which are derived by fitting the spectral energy distribution (SED) of the Herschel data.
Details of the data analysis are described by \cite{Sugitani2019}.
We regridded the maps of $N$(H$_2$) and $T_{\rm{dust}}$ 
onto the same grid
as that of the molecular data obtained by the 45 m telescope.
%\footnote{The observed intensity by Herschel data can be stated as
%%Equation 2%%%%%%%%%%
%\begin{equation}
%\label{eq:dust}
%$I_\nu=B_\nu(T_{\rm{d}})(1-e^{\tau_\nu}) \simeq B_\nu(T_{\rm{d}})\tau_\nu=B_\nu(T_{\rm{d}})\kappa_\nu \Sigma$,
%\end{equation}
%%%%%%%%%%%%%%%%%
%where $I_\nu$ is the observed intensity at a frequency $\nu$, $B_\nu(T_{\rm{d}})$ is
%the blackbody brightness of the object as a function of color temperature
%$T_{\rm{d}}$ , and $\tau_\nu$ is the source optical depth.
%We assume the dust opacity per unit mass of $\kappa_\nu = 0.1( \nu /1000.0 \times 10^{9})^{\beta}$ cm$^{2}$/g, where ${\beta}=2.0$}. 

%%%%%%%%%%%%%%%%%%%%%%%%%%%%%%%%%%%
%                                         RESULTS                                                 %
%%%%%%%%%%%%%%%%%%%%%%%%%%%%%%%%%%%

%\section{Results}
\section{Results}

%%%%%%%%%%%%%%%%%%%
% Spatial distributions of M17SWex %
%%%%%%%%%%%%%%%%%%%

\subsection{Spatial distributions of M17 SWex }

\begin{figure*}
\begin{center}
\includegraphics[scale=.2]{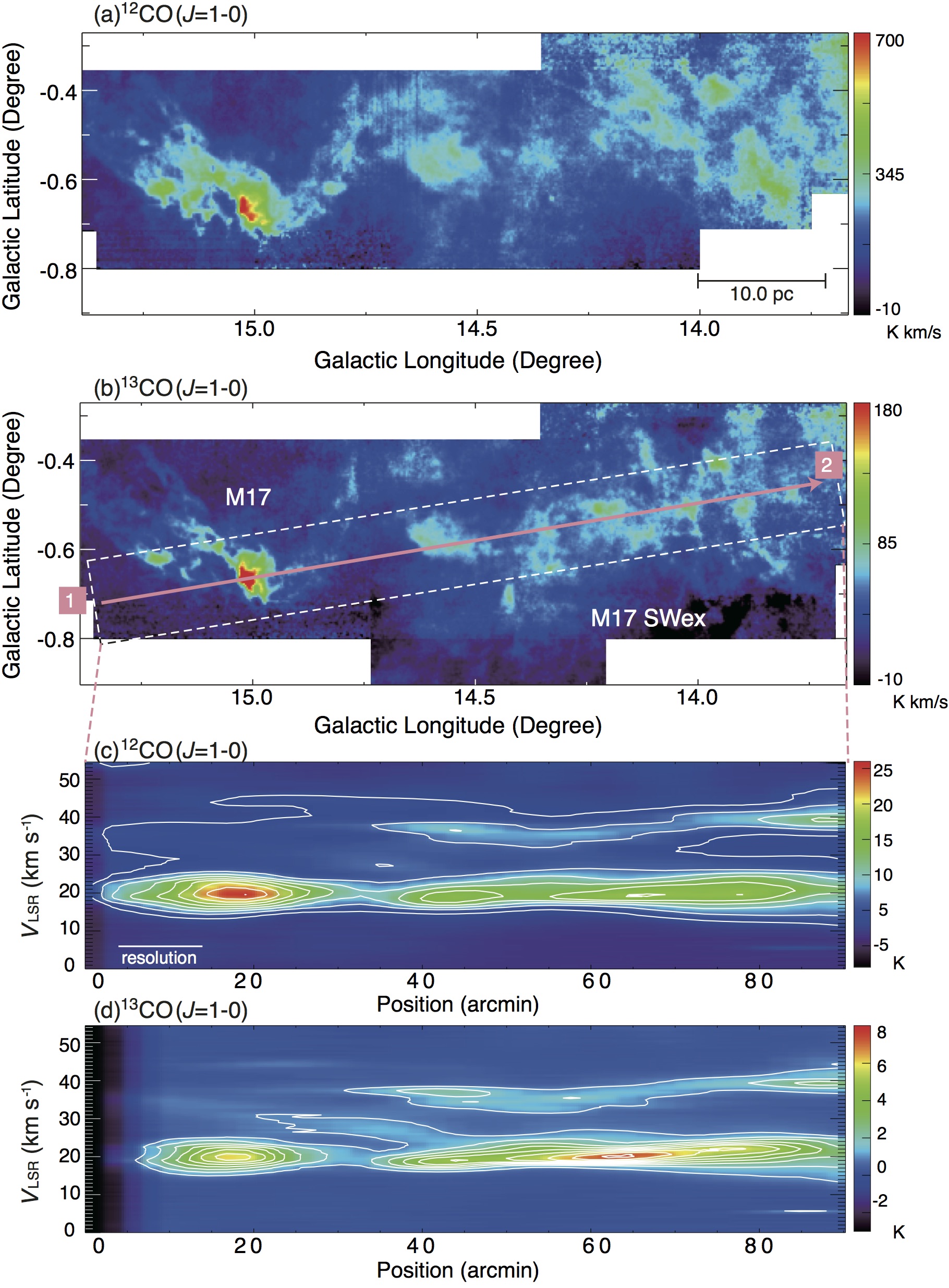}
\end{center}
\caption{
Integrated intensity maps of the (a) $^{12}$CO and (b) $^{13}$CO emission lines.
The velocity ranges used for the integration is $-10.0 \lesssim V_{\rm LSR} \lesssim 55.0 $ km s$^{-1}$
for both of the emission lines.
Position-velocity diagrams of the (c) $^{12}$CO and (d) $^{13}$CO emission lines
taken along the cut (1)-(2) in panel (b).
The lowest contour and the contour interval are both 3.0 K for panel (c)
and are both 1.0 K for panel (d).
As indicated in panel (b) by the white broken line, the angular resolution of the position-velocity diagrams
is set to $10\arcmin$ to cover most of the cores identified in this study.
The velocity resolution is set to 0.5 km s$^{-1}$.
\label{fig:PV}}
\end{figure*}

Figure \ref{fig:PV} shows the integrated intensity maps of the $^{12}$CO and $^{13}$CO emission lines.
As we show in panels (c) and (d) of the figure,
we made position-velocity (PV) diagrams across the M17 H{$\,${\sc ii}} region and M17 SWex 
along the line in panel (b).
As seen in the figure, the intense emission lines around $V_{\rm{LSR}}=20$ km s$^{-1}$
smoothly change from the M17 H{$\,${\sc ii}} region to M17 SWex in terms of the radial velocity, line width,
and the brightness temperature. We therefore assume that M17 SWex is physically connected with
the M17 H{$\,${\sc ii}} region.

\begin{figure*}
\begin{center}
\includegraphics[scale=.2]{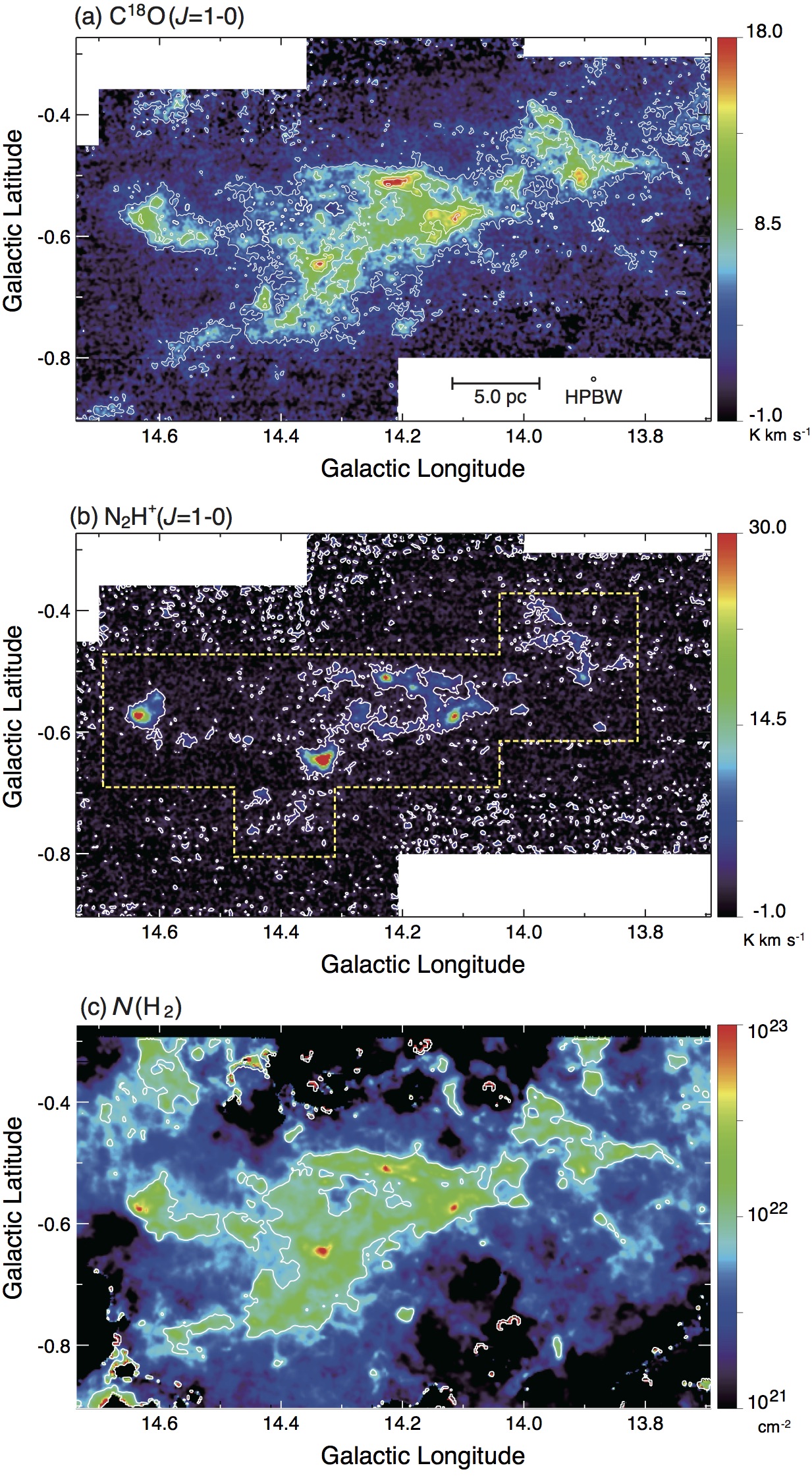}
\end{center}
\caption{
Integrated intensity maps of the (a) C$^{18}$O and (b) N$_2$H$^{+}$ emission lines.
The velocity ranges used for the integration are (a) $15.0 \lesssim V_{\rm LSR} \lesssim 26.0 $ km s$^{-1}$ and (b) $6.0 \lesssim V_{\rm LSR} \lesssim 34.0 $ km s$^{-1}$. 
(c) The H$_2$ column density map derived from the Herschel data. 
The lowest contour and the contour interval for the C$^{18}$O map are 4.0 K km s$^{-1}$. 
Contours are drawn at 2.8 K km$^{-1}$ and $1.0 \times 10^{22}$ cm$^{-2}$ in the N$_2$H$^{+}$ and $N$(H$_2$) maps, respectively.
\label{fig:iimap}}
\end{figure*}

Figure \ref{fig:iimap} shows the integrated intensity maps of the C$^{18}$O and N$_2$H$^{+}$ emission lines.
CCS emission was not detected at the present sensitivity.
The N$_2$H$^{+}$ and C$^{18}$O emission lines are distributed in the velocity range $6.0\lesssim V_{\rm{LSR}}\lesssim34.0$ km s$^{-1}$ and $15.0\lesssim V_{\rm{LSR}}\lesssim26.0$ km s$^{-1}$, respectively.
The map of C$^{18}$O traces well the shape of the flying dragon of the infrared dark cloud.
Since the signal-to-noise ratio of the N$_2$H$^{+}$ map is poor on the edge, 
we analyzed the data within the area surrounded by the yellow broken line shown on the map for N$_2$H$^{+}$ in this study.
For comparison, we also show the $N$(H$_2$) map calculated using the Herschel data by Sugitani et al. (2019)
 in figure \ref{fig:iimap}(c).
As seen in the panels (a) and (c), the distributions of C$^{18}$O and $N$(H$_2$) show a good correlation.
We also found that N$_2$H$^{+}$ is distributed in regions where the $N$(H$_2$) density is relatively high ($\gtsim10^{22}$cm$^{-2}$).

\begin{figure*}
\begin{center}
\includegraphics[scale=.2]{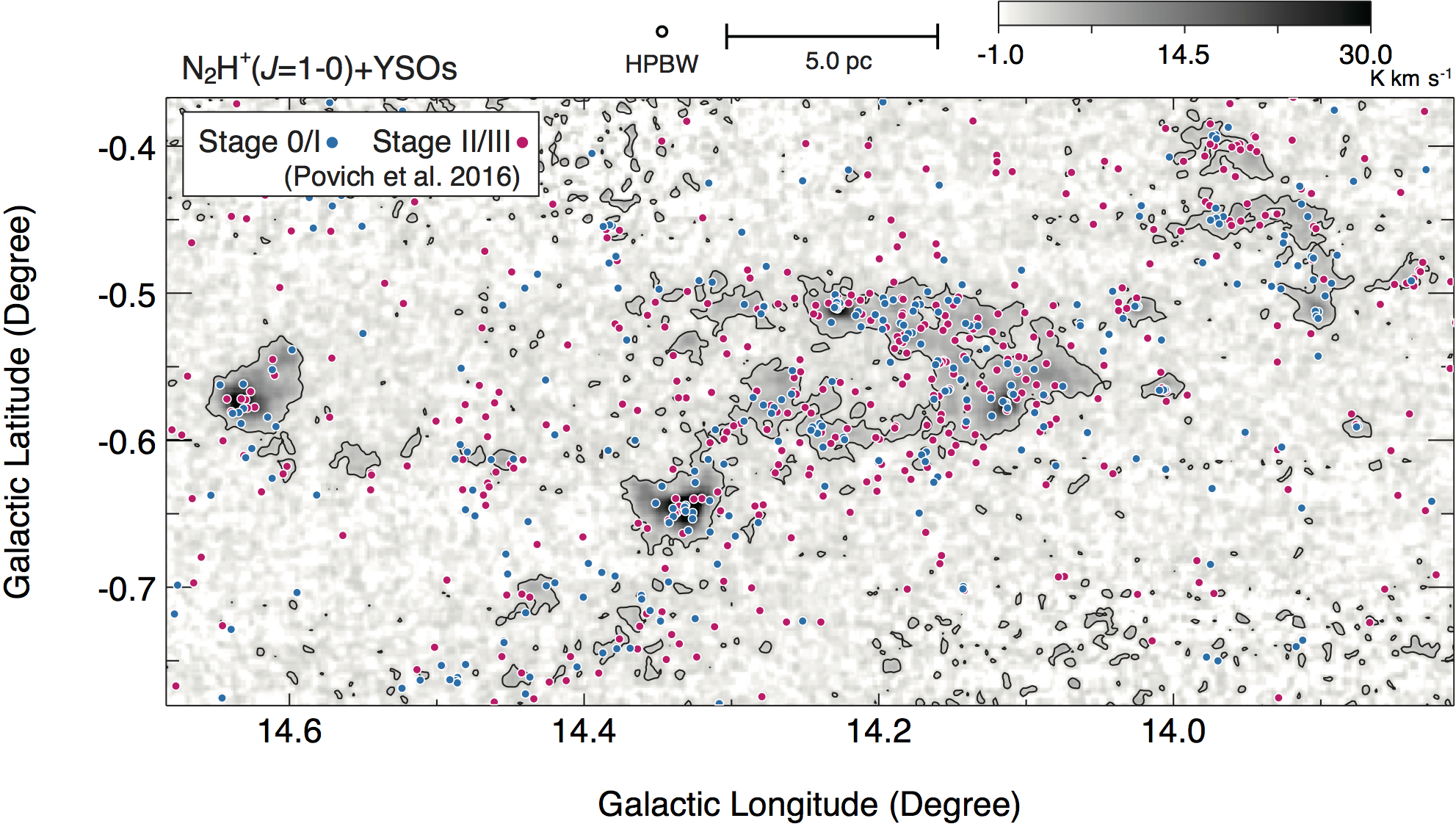}
\end{center}
\caption{
Distribution of YSOs associated with the M17 SWex region reported by \cite{Povich2016}. 
The contours denote the N$_2$H$^{+}$ integrated intensity map same as in figure \ref{fig:iimap} (b). 
Stage 0/I is the YSO accompanied by an infalling envelope, Stage II is the YSO with an optically thick circumstellar disk, 
Stage III is the YSO with an optically thin disk \cite[see][]{Povich2010}.
\label{fig:YSOs}}
\end{figure*}

\cite{Povich2010} made a YSOs survey around M17 SWex.
In the catalog they compiled, a total of 488 YSOs classified as Stage 0/I (YSOs accompanied by an infalling envelope), Stage II (YSOs with optically thick circumstellar disk), 
Stage III (YSOs with optically thin disk), and ``Ambiguous stage" are listed. 
\cite{Povich2016} updated the catalog. 
In the catalog, 840 stars are classified as Stage 0/I, Stage II/III, and Ambiguous stage.
We show the distributions of these YSOs associated with the observed region using the catalog of \cite{Povich2016} in figure \ref{fig:YSOs}.
We found that the Stage 0/I YSOs are distributed in dense parts of the N$_2$H$^{+}$ emission whereas the Stage II and III YSOs are distributed more randomly throughout the M17 SWex region.

%In the following, because the physical parameters of the YSOs are cataloged only in the results by \cite{Povich2010}, 
%we use their catalog to investigate the relation with physical parameters of the molecular emission lines.

%%%%%%%%%%%%%
% Identification of cores 
%%%%%%%%%%%%%

\subsection{Identification of cores}\label{sec:identification_of_cores}
In order to identify cores based on the N$_2$H$^{+}$ emission,
we apply the ``dendrogram'' algorithm\footnote{http://www.dendrograms.org} \cite[e.g.,][]{Rosolowsky}
to the N$_2$H$^{+}$ integrated intensity map.
%To compute the dendrogram, we have used the PYTHON implementation astrodendro.7
The dendrogram identifies three hierarchical structures of ``leaf", ``branch", and ``trunk".
We chose the minimum threshold intensity required to identify a parent tree structure to be $4\, \sigma$ (=2.8 K km s$^{-1}$) and a splitting threshold intensity required to identify structures to be $2\, \sigma$.
We only consider leaves with at least 25 pixels which are the equivalent area of the synthesized beam ($0.39$ arcmin$^{2}$), and we refer to them as ``cores".
As a result, a total of 46 cores are identified. 
We note that the results of the above core identification, e.g., the number and position of the cores, 
do not change significantly with other algorithms such as ``clumpfind''.

\begin{figure*}
\begin{center}
\includegraphics[scale=.4]{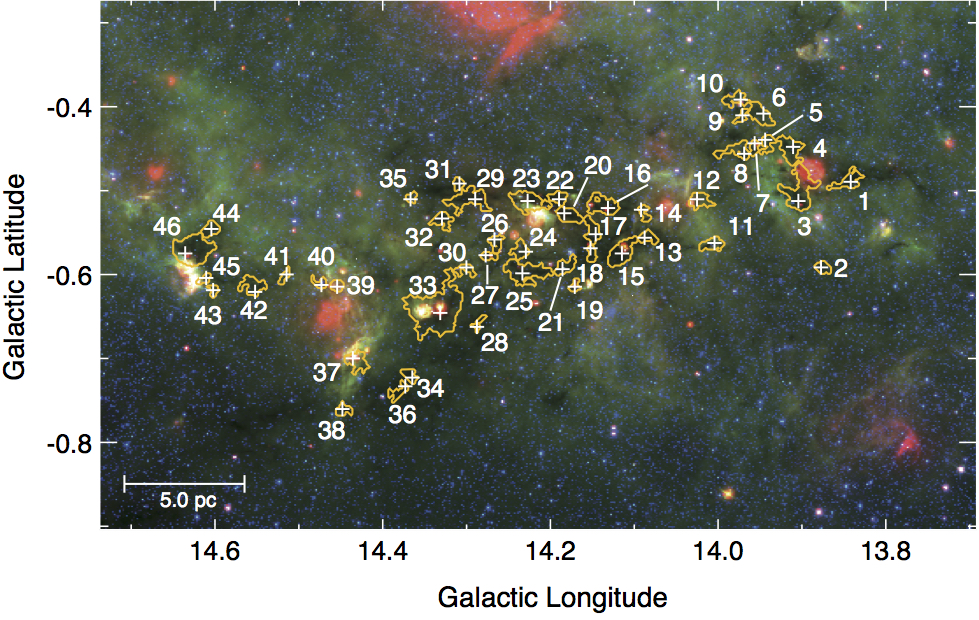}
\end{center}
\caption{
Distribution of the identified cores. 
The background is a $\it{Spitzer}$ composite image made with the 
$24.0 \micron$ (red), $8.0 \micron$ (green) and $3.6 \micron$ (blue) data.
The identified cores are numbered in the order of galactic longitude. 
The white plus sign is the peak position of the N$_2$H$^{+}$ integrated intensity of each core.
Orange contours denote the surface area of the cores defined at $2.8$ K kms $^{-1}$ ($=4 \sigma$).
\label{fig:core}}
\end{figure*}

Figure \ref{fig:core} shows the cores identified. 
We numbered the identified cores from No.1 to No.46 in the order of galactic longitude.
We searched for YSOs in the catalog of \cite{Povich2016} within the boundary of the cores 
and found that there are 40 cores with YSOs and 6 cores without  YSOs.
We assume that the YSOs are associated with the cores.
The peak position of the cores and the number of YSOs associated with each core are summarized in table  \ref{tab:core}.
%, the total mass of YSOs, and the total luminosity of YSOs associated with each core are summarized in table  \ref{tab:core}.
We found that four cores (Nos. 15, 23, 33, and 46) are especially large having a size of $\sim1$ pc.
These cores are accompanied by many YSOs. 
%and a mass of $>1000M_{\odot}$.
Among these, two cores correspond to the regions known as hub-N (=No. 33) and hub-S (=No. 15).

The distances to some of the detected cores can be found in the literature,
but there is an uncomfortable mismatch.
\cite{Wu2014} determined the distance of G$14.63-0.57$ (= core No.46) to be $1.83 ^{+0.08}_{-0.07}$ kpc
by a parallax measurement. On the other hand, \cite{Sato2010} reported a distance of $1.12\pm0.13$ kpc
for G$14.33-0.64$(= core No. 33) which is located not far from G14.63-0.57 on the plane of the sky. 
\cite{Povich2016} suggested that G$14.63-0.57$ is actually much closer than the other core.
However, as can be seen in figure \ref{fig:PV}, the radial velocities, line widths, and intensities of the observed
$^{12}$CO and $^{13}$CO lines change smoothly over the entire M17 SWex region including
cores No. 33 and 46, suggesting that they may belong to the same system.
In this paper, we assume that all of the cores identified in this work are located at 2.0 kpc \citep{Xu2011} for simplicity.
Parameters of the cores derived in this paper should be rescaled when their distances
are established precisely.
Complexity of the distance measurements of the
M17 SWex region in the literature is summarized and discussed by \cite{Povich2016} in detail.

\subsection{Physical parameters derived from the molecular line data }

Assuming local thermodynamic equilibrium (LTE),
%and that the observed emission lines are optically thin, 
we estimated the physical parameters of the cores.

In order to derive the excitation temperature based on the $^{12}$CO data 
as well as to derive the column density of C$^{18}$O,
we followed the method described by \cite{Shimoikura2018}.
Using the optically thick $^{12}$CO line ($\tau\gg1$), 
we estimated the excitation temperature $T^{\rm{co}}_{\rm ex}$ at the individual positions in the observed region.
The $^{12}$CO emission is widely distributed over the whole region in the observed velocity range (see figure \ref{fig:PV}).
The peak intensity of the $^{12}$CO spectra was measured for M17 SWex 
in the velocity range from 0 to 30 km s$^{-1}$, 
and we determined $T^{\rm{co}}_{\rm ex}$ by measuring the peak brightness temperature of the $^{12}$CO line in this velocity range.
Finally, the C$^{18}$O column density $N$(C$^{18}$O) was 
calculated from the C$^{18}$O intensity. 

Next, we derived the physical parameters of the N$_2$H$^{+}$ molecular emission line.
The procedure for the derivation is similar to that described by \cite{Shimoikura2019}.
We fitted the seven hyperfine components with multiple Gaussian functions under the assumption 
that all of the seven hyperfine lines have a single line width and excitation temperature.
The fitting was performed to the spectra within the area where the cores were identified.
We then derived the peak temperature $T_{\rm{mb}}$, the centroid velocity $V_{\rm{LSR}}$, the line width $\Delta V$, 
the excitation temperature $T^{\rm{N{_2}H^{+}}}_{\rm ex}$, and the total optical depth $\tau_{\rm{tot}}$ 
for all of the hyperfine components.
When the emission line is optically thin,
$\tau_{\rm{tot}}$ and $T^{\rm{N{_2}H^{+}}}_{\rm ex}$ cannot be estimated well simultaneously by the fitting. 
For such spectra, we estimated $T^{\rm{N{_2}H^{+}}}_{\rm ex}$ 
by interpolating $T^{\rm{N{_2}H^{+}}}_{\rm ex}$ at the neighboring positions 
with a weighted two-dimensional Gaussian function.
We then estimated the N$_2$H$^{+}$ column density $N$(N$_2$H$^{+}$) using the line parameters derived above
\cite[see][]{Shimoikura2019}. 

\begin{figure*}
\begin{center}
\includegraphics[scale=.4]{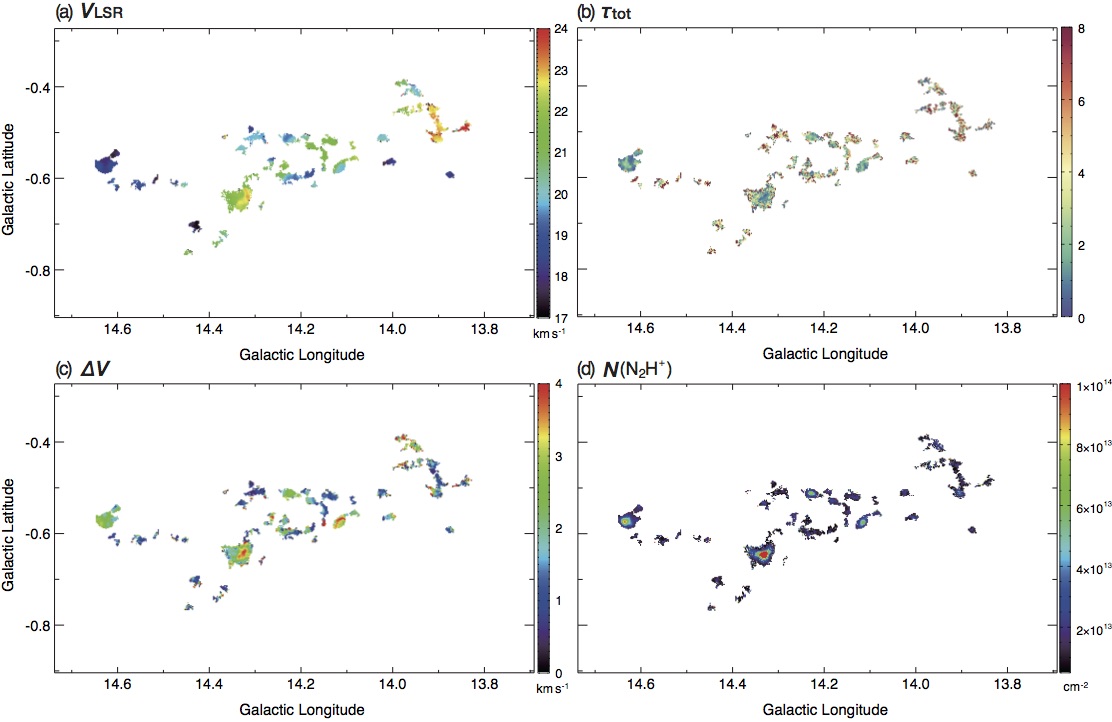}
\end{center}
\caption{
(a) The centroid velocity map, 
(b) total optical depth map,  
(c) line width map, and  
(d) column density map of the N$_2$H$^{+}$ emission line. 
These maps are obtained based on the hyperfine spectra fitting of the N$_2$H$^{+}$ data.
\label{fig:fit}}
\end{figure*}

\begin{figure*}
\begin{center}
\includegraphics[scale=.4]{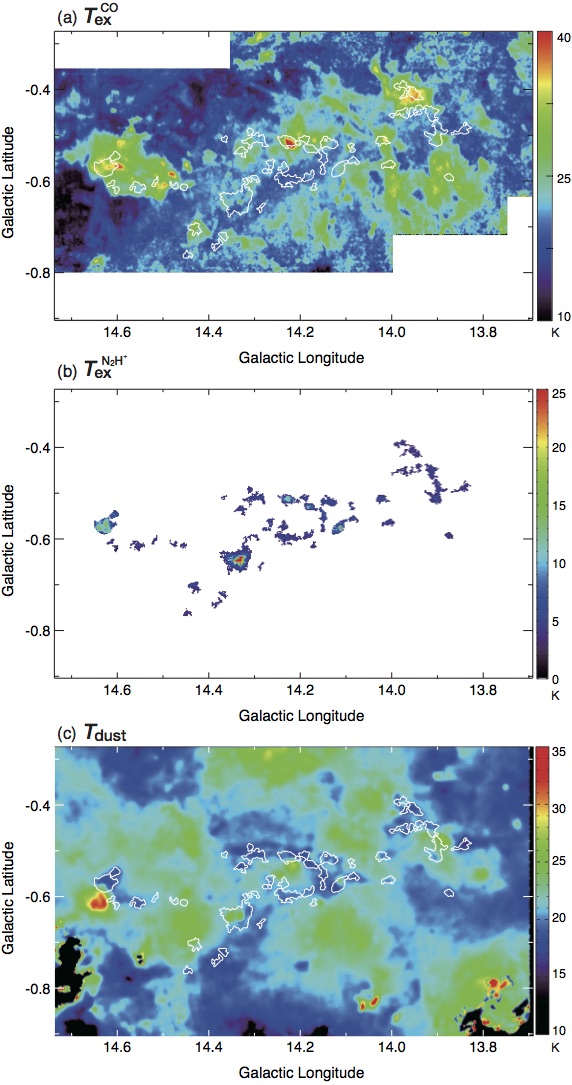}
\end{center}
\caption{
(a) Excitation temperature map derived from the $^{12}$CO data. 
(b) Excitation temperature map derived from the N$_2$H$^{+}$ data. 
(c) Dust temperature map derived from the Herschel data.
The cores shown in figure \ref{fig:core} are overlaid in panels (a) and (c) by the white contours.
\label{fig:temperature}}
\end{figure*}

In figure \ref{fig:fit}, we present each parameter map obtained from the hyperfine spectra fitting.
The line parameters of the peak positions of the cores are summarized in table \ref{tab:line_n2hp}.
The distributions of the temperatures $T^{\rm{co}}_{\rm ex}$, $T^{\rm{N{_2}H^{+}}}_{\rm ex}$, and $T_{\rm{dust}}$ 
are shown in figure \ref{fig:temperature}.

For each core, we measured the surface area $S$ at the 2.8 K km s$^{-1}$ (=$4\, \sigma$) 
contour level shown in figure \ref{fig:core},
and the radius $R_{\rm core}$ defined as $R_{\rm core}=\sqrt{S/\pi}$.
Next, using the $N(\rm{H}_2)$ data obtained from the optically thin dust data by Hershel, 
we estimated the mass of the core $M_{\rm{core}}$ as 
%%Equation 2%%%%%%%%%%
\begin{equation}
\label{eq:mass}
M_{\rm{core}}=\mu m_{\rm{H}} S \times \Sigma N(\rm{H}_2), 
\end{equation}
%%%%%%%%%%%%%%%%%
where $\mu$ is the mean molecular weight (2.8) and $m_{\rm{H}}$ is the hydrogen mass.
%, $\Omega$ is a surface area for 1 pixel ($7.5 \times 7.5$ arcsec$^{2}$).

We assume that the cores are a sphere of radius $R_{\rm core}$ with uniform density and temperature.
When the cores are dynamically stable, 
the virial theorem \citep{Spitzer} can be written as
%%Equation 3%%%%%%%%%%
\begin{equation}
\label{eq:Pext}
4\pi {R_{\rm core}}^3 P_{\rm ext}=
\frac{3kT}{\mu m_{\rm H}}
{M_{\rm core}}-
\frac{3G}{5R_{\rm core}}
{M^2_{\rm core}}\ ,
\end{equation} 
%%%%%%%%%%%%%%%%%
where $P_{\rm ext}$ is the external pressure by the surrounding gas, $k$ is the Boltzmann constant,
and $G$ is the gravitational constant.
We assume temperature of the cores $T$ can be expressed as $T={\mu m_{\rm H}}\Delta V^2/8k(\rm{ln}2)$
which is equivalent to the Doppler temperature of the gas.
We then estimated the virial mass $M_{\rm vir}$ at $P_{\rm ext}$=0 using
%%Equation 2%%%%%%%%%%
\begin{equation}
\label{eq:M_Vir}
M_{\rm vir}=
\frac{5R}{G}
\frac{{\Delta V}^{2}}{8\,{\rm ln}2}\ ,
\end{equation} 
%%%%%%%%%%%%%%%%%
where ${\Delta V}$ is the line width obtained by the hyperfine fitting at the peak position of the cores. 
We list the derived parameters in table \ref{tab:mass}.

Furthermore, the fractional abundances of C$^{18}$O and N$_2$H$^{+}$, $f$(C$^{18}$O) and $f$(N$_2$H$^{+}$), respectively,
are directly derived from the ratios of each column density and $N$(H$_2$).  

The constants of the observed molecular lines used to derive the parameters in this section are summarized by 
\citet[][see their table 2]{Shimoikura2018}.
We found that the derived parameters are in the ranges
$0.2 \lesssim R_{\rm core} \lesssim 1.0$ pc,
$43 \lesssim M_{\rm core} \lesssim 3026$ ${M}_{\odot}$,
$3.00 \lesssim T^{\rm{N{_2}H^{+}}}_{\rm ex} \lesssim 31.68$ K, and
$0.76 \lesssim \Delta V \lesssim 4.00$ km s$^{-1}$.
Among the 46 cores, we found the four massive cores (Nos. 15, 23, 33, and 46) with a mass of more than $1000 M_{\odot}$.
These cores have larger $\Delta V$ ($\gtsim2.5$ km s$^{-1}$) and higher $T^{\rm{N{_2}H^{+}}}_{\rm ex}$ ($\gtsim 10$ K) than the other cores.
$T^{\rm{N{_2}H^{+}}}_{\rm ex}$and $\Delta V$ of the other cores are
$<10$ K and $<2$ km s$^{-1}$, respectively, indicating that
%and of the other cores are $<10$ K, and those of $\Delta V$ is mostly $<2$ km s$^{-1}$.
%These indicate that
$T^{\rm{N{_2}H^{+}}}_{\rm ex}$ and $\Delta V$ increase as $M_{\rm core}$ increases. 

\begin{figure*}
\begin{center}
\includegraphics[scale=.3]{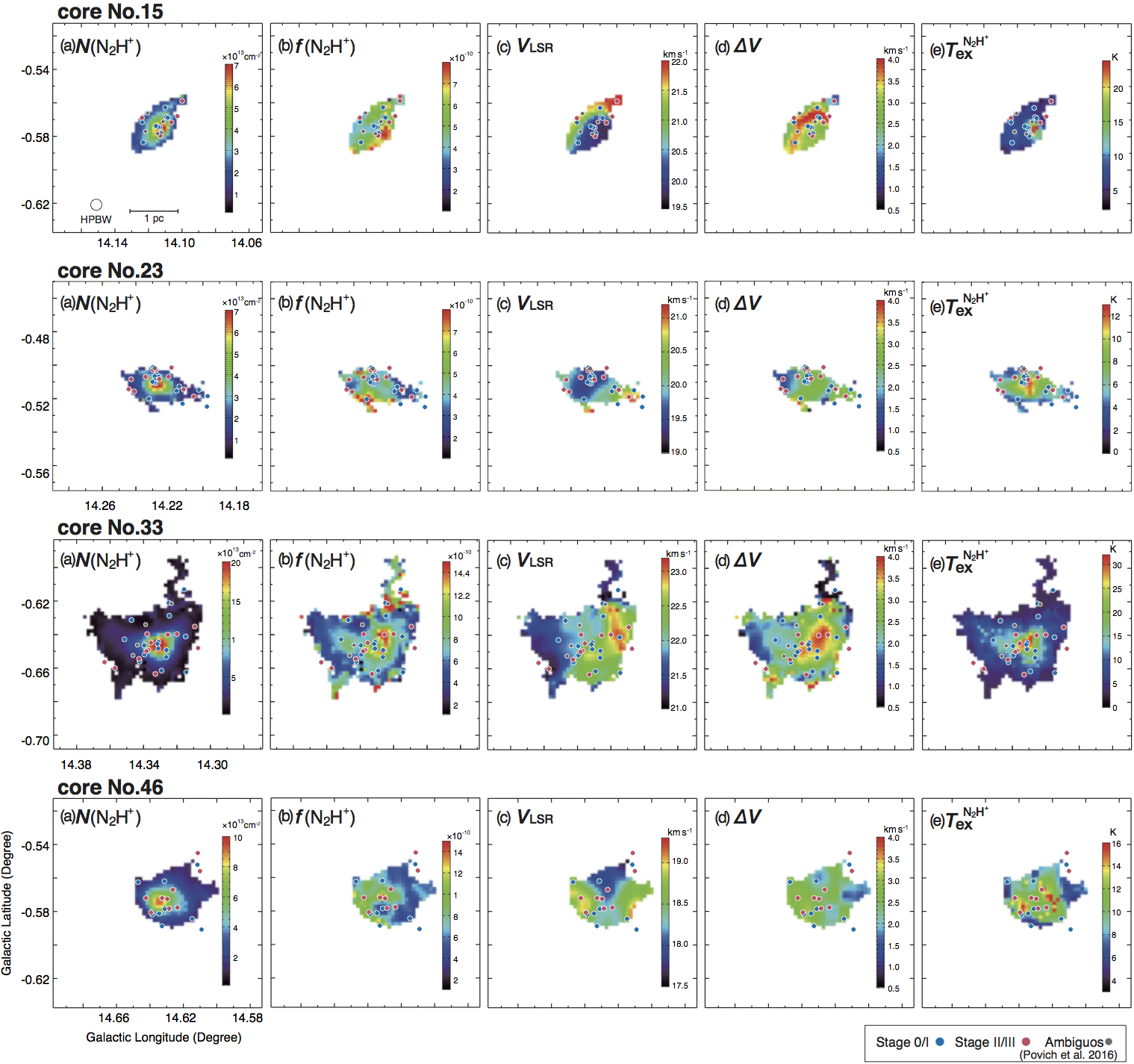}
\end{center}
\caption{
The N${_2}$H$^{+}$ parameter maps for cores No. 15, 23, 33, and 46.
(a) The $N$(N$_2$H$^{+}$) map.
(b) The $f$(N$_2$H$^{+}$) map.
(c) The $V_{\rm LSR}$ map.
(d) The $\Delta V$ map.
(e) The $T^{\rm{N{_2}H^{+}}}_{\rm ex}$ map.
\label{fig:4core}}
\end{figure*}

\cite{Busquet} revealed that the velocity dispersion is locally enhanced ($\sigma\sim1$ km s$^{-1}$) toward hub-N (No.23) and hub-S (No.15).
Based on the SMA images, \cite{Busquet2016} show that the two hubs fragment into several dust condensations. 
The large $\Delta V$ of the four cores suggest that
there are multiple unresolved sub-cores not only in cores No. 15 and 23 but also in cores No. 33 and 46. 
In figure \ref{fig:4core}, we show the parameter maps for the four cores.
It seems that there is a correlation between the distributions of $T^{\rm{N{_2}H^{+}}}_{\rm ex}$ and those of YSOs. 
In addition, it appears that there is an anti-correlation between the distributions of $f$(N$_2$H$^{+}$) and those of YSOs. 
For the relationships between the distribution of $\Delta V$ and those of YSOs, the correlation is not clear.

%%%%%%%%%%%%%%%%%%%
% Correlation between N (N 2 H +) and N (H 2)  %
%%%%%%%%%%%%%%%%%%%
\subsection{Correlation between $N$(N$_2$H$^{+}$) and $N$(H$_2$) for the cores}

To estimate the total masses of cores from $N$(N$_2$H$^{+}$),
abundance ratio of $f$(N$_2$H$^{+}$)$=3\times10^{-10}$ reported by
\cite{Caselli1995} has widely been used in star-forming regions.
However, as seen in figure \ref{fig:4core}, $f$(N$_2$H$^{+}$) varies
even within a single core.
In this study, we investigate the relationship between $N$(N$_2$H$^{+}$) and $N$(H$_2$)
of the identified cores.

\begin{figure*}
\begin{center}
\includegraphics[scale=.4]{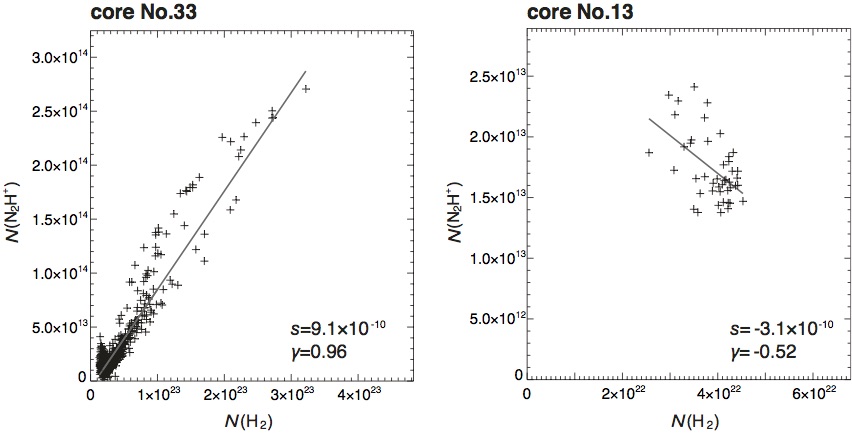}
\end{center}
\caption{
The relation between $N$(N$_2$H$^{+}$) and $N$(H$_2$) for
cores No.33 (left) and 13 (right).
\label{fig:core33}}
\end{figure*}

For each of the cores, we made a plot of $N$(N$_2$H$^{+}$) vs. $N$(H$_2$), and
found that $N$(N$_2$H$^{+}$) correlates roughly linearly with $N$(H$_2$) in most cases.
We show two examples in figure \ref{fig:core33}.
We performed a linear least-square fit to the plots as,
%%Equation 2%%%%%%%%%%
\begin{equation}
\label{eq:X}
%N({\rm{H}}_2)=X N(\rm{N}_2H^{+})+\it{i}.
N(\rm{N}_2H^{+})=\it{s N}({\rm{H}_2})+\it{i}
\end{equation} 
%%%%%%%%%%%%%%%%% 
where $s$ and $i$ are the slope and intercept of the linear relation, respectively.
In general, $N$(N$_2$H$^+$) represents the column density only of the limited region
denser than the critical density to excite the N$_2$H$^+$ line, but $N$(H$_2$) derived
from the dust emission traces the total column density along the line-of-sight including the diffuse surroundings.
Therefore,  $i$ in the above can be used to remove the contribution from the low density regions not emitting
the N$_2$H$^+$ line, and $s$ should better trace the average fractional abundance of N$_2$H$^+$ in the cores.

In table \ref{tab:fractional}, we summarize the obtained values of $s$ and $i$
together with the correlation coefficients $\gamma$.
As listed in the table, there is a large variation in $s$,
and its median value is $s=3.5\times10^{-10}$
which is consistent with those measured in other clouds
 \citep[e.g.,][]{Tanaka2013, Shimoikura2019}.

Among the detected 46 cores, there are 15 cores showing a good correlation with
$\gamma \gtsim0.6$, and 14 cores showing almost no correlation ($\gamma \ltsim0.2$).
Most of the cores have a positive value of $s$($>0$), but there are some cores having negative 
values ($s<0$, e.g., core No. 13), suggesting that N$_2$H$^{+}$ is probably adsorbed onto dust in the densest
part of the cores \cite[e.g.,][]{Bergin2002,Belloche}.

%%%%%%%%%%%%%%%%%%%
% Correlation between gas and YSOs %
%%%%%%%%%%%%%%%%%%%
\subsection{Relationship between the cores and YSOs}

We attempt to investigate the relationship between the mass of the cores  
and those of YSOs forming in the cores.
For this, we estimated the total stellar mass $\Sigma M_{\star}$ associated with each core.
%using the catalog of YSOs made by \cite{Povich2010}.
The catalogue of \cite{Povich2010} contains a number of good candidate YSOs, but some of the
estimated parameters such as the stellar mass could be erroneous due to
ambiguities of the adopted model.
We therefore made use of the newer catalog by \citet{Povich2016}
which is more complete but does not present stellar masses,
and we adopted the following rough method to estimate $\Sigma M_{\star}$:
First, we counted the catalogued YSOs located within the extent of each core,
and assumed that the median value of their masses is $M_{\rm medi}=3$ ${M}_{\odot}$
(Povich 2019, private communication).
%half of the YSOs has a mass of greater than the certain value $M_{\rm th}=3$ ${M}_{\odot}$.
Second, we assumed the stellar initial mass function (IMF) by \cite{Kroupa}
and scaled the IMF to match the observed number of YSOs having a mass greater than
$M_{\rm medi}$ in each core.
Finally, we estimated $\Sigma M_{\star}$ in the cores by integrating the scaled IMF.
Resulting values of $\Sigma M_{\star}$ are summarized in the last column of table \ref{tab:mass}.
The total stellar mass in the table amounts to $\sim2900$ ${M}_{\odot}$.
%(If we adopt $M_{\rm medi}=2$ ${M}_{\odot}$ instead of $3$ ${M}_{\odot}$,
%the total stellar mass will be rescaled to $\sim1700$ ${M}_{\odot}$.)

\begin{figure*}
\begin{center}
\includegraphics[scale=.4]{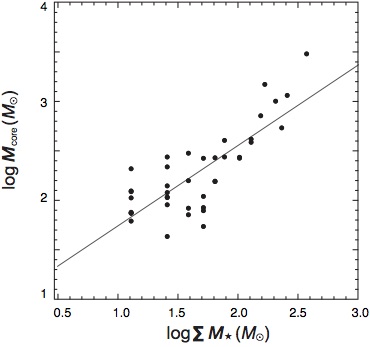}
\end{center}
\caption{
The core mass $M_{\rm{core}}$ vs. the total stellar mass $\Sigma M_{\star}$.
The solid line is the least-squares fit given in eqation \ref{eq:star1}.
%(b) $M_{\rm{core}}$ vs. the total stellar luminosity $\Sigma L_{\star}$.
%The solid line is the least-squares fit given in equation \ref{eq:star2}.
%(c) The total accretion rate $\Sigma \dot{M}_{\rm{env}}$ vs. the momentum of the core.
%The solid line is the least-squares fit given in Eq(1).
\label{fig:core_peak}}
\end{figure*}

In figure \ref{fig:core_peak}, we display the relation of $M_{\rm{core}}$ vs. $\Sigma M_{\star}$
%(for $M_{\rm medi}=3$ ${M}_{\odot}$)
in logarithmic scale.
They are roughly correlated, and the relation can be fitted as
%%Equation 3+4%%%%%%%%%%
\begin{equation}
\label{eq:star1}
{\rm log}\ M_{\rm core}=(0.93\pm0.17)+(0.81\pm0.10){\rm log}\ \Sigma M_{\star}\ 
\end{equation}
%%%%%%%%%%%%%%%%%
where $M_{\rm core}$ and $\Sigma M_{\star}$ are in units of $M_\odot$.
The index ($0.81$) is close to unity, indicating that the total stellar mass
is roughly proportional to the natal core mass.

%%%%%%%%%%%%%%%%%%%
% SFE
%%%%%%%%%%%%%%%%%%%

\subsection{Star Formation Efficiency of M17 SWex}\label{sec:SFE}

\begin{figure*}
\begin{center}
\includegraphics[scale=.4]{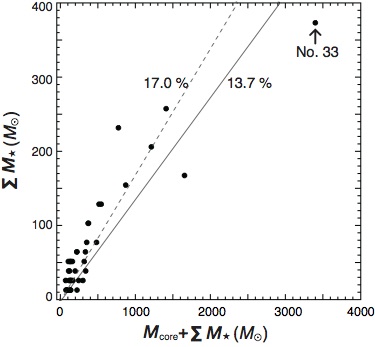}
\end{center}
\caption{
The total stellar mass of YSOs associated with each core $\Sigma M_{\star}$ vs. 
the total mass of the system for each core ($M_{\rm core} + \Sigma M_{\star}$). 
The linear least-square fit (SFE=$13.7\%$) is shown by the solid line.
The broken line denotes the best fit when we exclude the core No.33 from the fit.
%Black broken lines are for SFE = $10.0\%$ and $1.0\%$.
\label{fig:SFE}}
\end{figure*}

We show a plot of $\Sigma M_\star$ vs. $M_{\rm core}$+$\Sigma M_\star$ in figure \ref{fig:SFE}.
Star Formation Efficiency (SFE) is a fundamental parameter of cluster-formation process \citep{lada2003}
and is calculated as $\Sigma M_\star /(M_{\rm core}+\Sigma M_\star$ ).
We investigated the SFE of M17 SWex by fitting the all of the data points in figure \ref{fig:SFE}.
The result is shown in the figure by the solid line which corresponds to
SFE$=13.7\%$.
Many of the data points smaller than $M_{\rm core}+\Sigma M_\star=2000$ $M_\odot$
are above the solid line because of core No. 33,
whose distance might be very different from the other cores (see section \ref{sec:identification_of_cores}).
We would obtained SFE$=17.0\%$ if we exclude this core from the fit (the broken line in the figure).
The total SFE obtained from the total stellar mass ($\sim2920$ $M_\odot$) and
the total mass of the cores ($\sim14300$ $M_\odot$) is $16.9\%$.
% which is close to the latter SFE.

As seen in figure \ref{fig:YSOs}, there are YSOs outside the extents of the detected cores.
\cite{Povich2010} estimated $\Sigma M_\star$ for M17 SWex to be $8\times10^{3}M_{\odot}$ 
by integrating the IMF of \cite{Kroupa} over $M_{\star}\gtsim0.1M_{\odot}$.
We estimated the total gas mass of the entire M17 SWex region to be 
$7.8\times10^{4}M_{\odot}$
by integrating $N(\rm{H}_2)$ within the area defined by the $1.0\times10^{21}$ H$_2$ cm$^{-2}$ contour.
These values infer a average SFE of $9.3 \%$ for the entire M17 SWex region.

According to \cite{lada2003} who cataloged the SFEs for nearby embedded clusters, the SFEs range
from approximately 10 to 30 $\%$.
The values of $9.3\%$ and $\sim17\%$
obtained in M17 SWex is comparable to those
of the cluster-forming regions.
\section{Discussion}

\subsection{Dynamical state of the cores \label{sec:dynamical}}

\begin{figure*}
\begin{center}
\includegraphics[scale=.4]{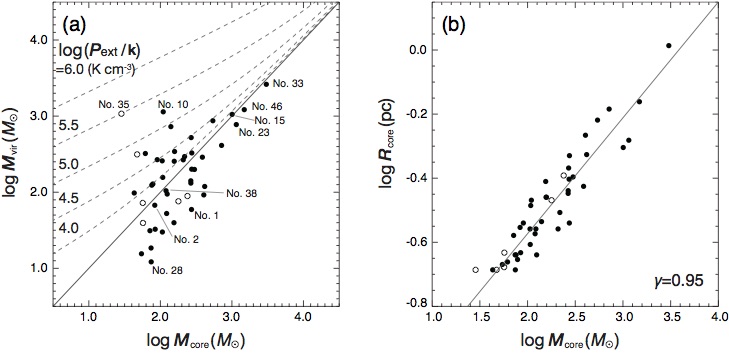}
\end{center}
\caption{
(a) Relationship between $M_{\rm vir}$ and $M_{\rm core}$.
Filled and open circles represent the cores with and without YSOs, respectively.
The solid line shows the external pressure $P_{\rm ext}$ = 0 where$M_{\rm core} = M_{\rm vir}$, 
and the dotted line show the virial masses in equilibrium with the external pressure log ($P_{\rm ext} / k$) = 4.0, 4.5, 5.0, 5.5, 6.0 K cm$^{-3}$.
(b) Relationship between $R_{\rm core}$ and $M_{\rm core}$. 
The black line shows the fitting result of the relation ($R_{\rm core}=0.05 M_{\rm core}^{0.37}$, see text).
\label{fig:Mvir-Mcore}}
\end{figure*}

We investigate the dynamical state of the cores.
Figure \ref{fig:Mvir-Mcore}(a) shows the relation between $M_{\rm{vir}}$ and $M_{\rm{core}}$.
The solid line indicates $M_{\rm{core}}=M_{\rm{vir}}$.
Focusing on the relationship between $M_{\rm{core}}$ and $M_{\rm{vir}}$, 
the observed cores can be divided into the following three types, and we discuss their characteristics.

\begin{enumerate}
\item We found that many cores satisfy the condition of $M_{\rm{core}}\simeq M_{\rm{vir}}$, 
indicating that the cores are dynamically stable if there is no external pressure ($P_{\rm ext}$=0).
More massive cores tend to be in the gravitational virial equilibrium.
%The other cores that satisfy this condition (e.g., cores No.2 and 38) are isolated from other cores and are considered to be in a relatively low density region where $P_{\rm ext}$ is low.

\item Cores satisfying the condition of $M_{\rm{core}}>M_{\rm{vir}}$ 
may be contracting because they cannot prevent collapsing by the internal pressure even when $P_{\rm ext}$=0.
The value of $\Delta V$ for such cores (e.g., cores No.1 and 28) is relatively small,
being less than 1 km s$^{-1}$ (table \ref{tab:line_n2hp}).
%in the order of 0.6 to 1.2 km s$^{-1}$. 
%The non-turbulent cores are located on the edge of the clump.

\item There are some cores under the condition of $M_{\rm{core}}<M_{\rm{vir}}$ (e.g., cores No.10 and 35), 
and such cores will disperse under the condition $P_{\rm ext}=0$.
However, they
can be
%are
confined in the pressure of the high density gas surrounding the cores.
We suggest that such cores 
are highly influenced by $P_{\rm ext}$ in equation \ref{eq:Pext}. 
Virial mass taking into account $P_{\rm ext}$ can be expressed as, 
%%Equation 4%%%%%%%%%%
\begin{equation}
\label{eq:Pext2}
{M_{\rm vir}}=
{M_{\rm core}}+
\frac{20\pi}{3G}
{\rm{A}}{M^{4{\rm B}-1}_{\rm core}} P_{\rm ext}\ ,
\end{equation} 
%%%%%%%%%%%%%%%%%
where A and B are constants to relate $M_{\rm core}$ and $R_{\rm{core}}$ as $R_{\rm{core}}= {\rm A} M^{\rm B}_{\rm core}$.
We derived A = 0.05 and B = 0.37 for $R_{\rm{core}}$ and $M_{\rm core}$ in units of pc and $M_{\odot}$, respectively,
by fitting the plot shown in figure \ref{fig:Mvir-Mcore}(b) with the relation.
In figure \ref{fig:Mvir-Mcore}(a), we show $M_{\rm{vir}}$ 
by the broken lines calculated based on equation \ref{eq:Pext2} for various $P_{\rm ext} (\neq0)$.
As seen in the figure, comparing the position of the plot with the lines, most of the cores
can be
%are considered to be
dynamically stable under $P_{\rm ext}/k=10^{5.0}-10^{5.5}$ K cm$^{-3}$.

\end{enumerate}

In summary, we suggest that many of the cores can be dynamically stable if we assume a large external pressure. 
For cores under the condition $M_{\rm{core}}>M_{\rm{vir}}$, 
the internal pressure due to turbulence is too small, and they should contract by the self-gravity if there is
no magnetic field strong enough to support them. We will further discuss this point in section \ref{sec:magneticfiled}.

%{\color{red}{\cite{Ohashi2016} suggested that the internal support of cores in the two hubs identified with the ALMA
%observations is insufficient to support the cores against the self-gravity and that the cores are undergoing dynamical collapse.}}

%The relatively small cores are surrounded by high density gas of about nH 2 =105 H 2 cm 3.
%If the temperature of the gas is about 10 K (Table 3.2), these cores are considered to be just under this external pressure.
%they appear to be largely pressure-confined rather than strongly gravitationally bound, and are not strongly centrally concentrated. 

%%%%%%%%%%%%%%%%%%%
% density structure
%%%%%%%%%%%%%%%%%%%
\subsection{Density structure of the cores}

We investigate the distribution of $N({\rm{H}_2})$ of the cores
as a function of the projected distance $R$ from the peak of $N({\rm{H}_2})$.
We found that profile of $N({\rm{H}_2})$ for the cores
are found to be different for each core.
In the case of a spherical core,
we assume that the number density of H$_2$ molecules $\rho$ is expressed as 
$\rho (r) \propto r^{-\alpha}$, where $r$ is the distance from the center of the sphere.
The physical state of the cores can be estimated by examining the value of $\alpha$.
For example, a density profile of gravitationally stable cores and free-falling cores
show $\alpha=2.0$ and 1.5, respectively \citep[e.g.,][]{Shu1977,Motte2001,Shimoikura2016}.

\begin{figure*}
\begin{center}
\includegraphics[scale=.4]{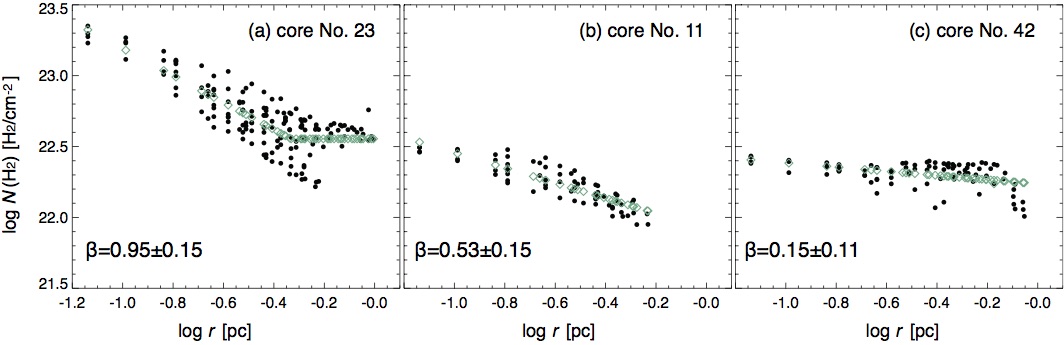}
\end{center}
\caption{
The distribution of $N({\rm{H}_2})$ as a function of the projected distance from the peak of $N({\rm{H}_2})$
for cores (a) No. 23, (b) No.11, and (c) No.42.
Green diamond represents a fit (see text).
\label{fig:dis}}
\end{figure*}

To investigate the density structure of the cores,
we fitted the relations between $N({\rm{H}_2})$ and $R$ of the cores with 
$N({\rm{H}_2})=N_{0} (R/R_{0})^{-\beta}+C$
where $N_0$, $R_0$, and $C$ are constants.
The obtained values of the index $\beta$ are roughly divided into three types:
(a) $0.7-1.2$ for the four massive cores,
(b) $0.3-0.6$ for the smaller cores with YSOs, 
and (c) $0.1-0.3$ for the cores without YSOs.
Because $\beta=\alpha-1$,
the values of $\beta$ infer $\alpha=1.7- 2.2, 1.3-1.6$, and $1.1-1.3$ 
for the cases (a), (b), and (c) in the above, respectively.
In figure \ref{fig:dis}, we show typical examples of the
$N$(H$_2$) vs. $R$ relations for the 3 types.
There are 4, 28, and 14 cores belonging the types (a), (b), and (c), respectively.

%We found that profile of $N({\rm{H}_2})$ for the cores
%are found to be different for each core, and $\beta$ are roughly divided into three types:
%(a) the four massive cores with a highest density contrast,
%(b) the smaller cores with YSOs with a small density contrast, 
%and (c) the cores without YSOs with a flat density contrast.
%Figure \ref{fig:dis} shows typical examples of the (a)-(c) types of the $N$(H$_2$) distribution 
%shown as a function of $R$ from the $N$(H$_2$) peak.
%The number of cores of each type is 4, 28, 14 respectively.

The results suggest that the cores of (a) and (b) types are likely to be gravitationally stable and free-falling, respectively.
%\cite{Povich2016} suggest that massive cores identified by the MALT90 survey \citep{Foster2011} 
%around hub-N (core No.23) and hub-S (core No.15) are still in the process of accreting mass.
%Our results are consistent with their suggestion.
Note that, in the case of the type (c) cores,  they show a flat density profile
and there is a possibility that the cores are not resolved at the present resolution due to their small size.

\subsection{Star formation in M17 SWex}\label{sec:magneticfiled}

As found in section \ref{sec:dynamical}, some of the cores satisfy the condition $M_{\rm{core}}>M_{\rm{vir}}$, 
which means that the cores are contracting if we ignore the effects of the magnetic field.
\cite{Ohashi2016} observed small cores in the two hubs identified with ALMA observations
which correspond to the cores No.12--23 in our work. 
They found that the small cores have viral parameters (defined as $M_{\rm{vir}}/M_{\rm core}$) in the range 0.1-1.3,
which is comparable to our results.
Based on the analysis of the virial parameters, 
\cite{Ohashi2016} speculated that mass accretion onto the cores via the hubs could produce massive stars in the future. 
However, their values of the virial parameters do not take into account the effect of the magnetic field.
In order to investigate whether the cores are really contracting or not,
a quantitative estimate of the magnetic field is needed.

Recently, \cite{Sugitani2019} conducted near-infrared polarization observations toward M17 SWex,
and revealed the overall distribution of the magnetic field around the individual filamentary structures.
They found that the magnetic field in M17 SWex is mostly perpendicular to the elongation of the filamentary clouds,
and they estimated the strength of the magnetic field to be $70-300$ $\mu$G using the
Davis-Chandrasekhar-Fermi method.
Based on these results, \cite{Sugitani2019} suggested 
that the clouds in M17 SWex are likely to be magnetically critical or sub-critical,
which may be responsible for the present low-level of massive star formation in M17 SWex.

The critical mass of the cores that can be supported by the magnetic field 
is expressed as $M_{\rm{cr}}/M_{\odot}\simeq2\times10^{2}(B/30\mu$G)($R_{\rm core}/1$pc)$^{2}$ \citep[e.g.,][]{Nakano,Shu}.
Using $R_{\rm core}$ and $M_{\rm core}$ in table \ref{tab:mass}, 
we found that the magnetic field of $\sim100-600$ $\mu$G is necessary for these cores to be magnetically critical.
The range of the magnetic field is consistent with or slightly larger than the values $70 - 300$ $\mu$G reported by
\citet{Sugitani2019}.
%{\color{blue}{However, the values do not take into account the components along the line of sight.
%In addition, the magnetic }} 
We should note that \citet{Sugitani2019} measured the magnetic field at the periphery of M17 SWex
where the background stars were well detected in the near-infrared, 
and thus, the magnetic field inside the dense cores can be
stronger than the values they obtained, if the magnetic field is frozen into the dense cores.
In addition, their values are for the component projected on the plane of the sky but
do not include the component parallel to the line-of-sight.
As \citet{Sugitani2019} suggested,
we conclude that the cores detected in this study are likely to be magnetically critical or sub-critical, 
and the magnetic field prevents the cores from collapsing, which may be the cause of the inactive massive
star formation in M17 SWex. 

However, it is rather mysterious that the total SFE of M17 SWex ($\sim 9-17\%$) is comparable to
those of other cluster-forming clumps producing massive stars (section \ref{sec:SFE}),
indicating that low-mass (and intermediate-mass) stars are forming in spite of the magnetic fileds.
We imagine that small cores producing the small stars, which may not be detected by our observations,
can lose the supporting force by the magnetic field soon via the ambipolar diffusion because of their small sizes. 
In addition, sporadic collisions of the internal substructures of the cores/clumps (such as ``fibers") may also induce
formation of low-mass stars via sudden increment of density \citep{Dobashi2014,Matsumoto2015}.
These effects can help the formation of low-mass stars.
For the massive cores we identified in M17 SWex, we suggest that they are unlikely to collapse at the moment
due to the support by the magnetic field, but they may eventually become supercritical to collapse and
form massive stars if they lose the clump-supporting force by the magnetic field, or if they gain more mass
via mass accretion from the surroundings.

%%%%%%%%%%%%%%%%%%%%%%%%%%%%%%%%%%%
%                                         Conclusions                                              %
%%%%%%%%%%%%%%%%%%%%%%%%%%%%%%%%%%%

\section{ CONCLUSIONS }\label{sec:conclusions}
An N$_2$H$^{+}$ survey for molecular cores was made toward the M17 southwest extension (M17 SWex) region. 
We identified 46 molecular cores.
Among the 46 cores, 40 of them are associated with Young Stellar Objects (YSOs) and the
other 6 cores are not associated with any YSOs.
Investigation of the core properties was also done by using the $^{12}$CO, $^{13}$CO, and C$^{18}$O data
as well as $N$(H$_2$) the column density of H$_2$ and the dust temperature derived from Herschel data.
We also used physical properties of YSOs found in the literature.

We summarize the main findings of this study as follows:

\begin{enumerate}
\item
We derived the masses of the individual cores from the $N$(H$_2$) data 
assuming a distance of 2.0 kpc, 
and found that the core masses vary in the range from 43$M_{\odot}$ to 3026$M_{\odot}$.
Based on the N$_2$H$^{+}$ data, we derived the physical parameters of the cores
and presented each parameter map.
The derived excitation temperature and the line width range
from 3.0 K to 31.7 K and from 0.8 km s$^{-1}$ to 4.0 km s$^{-1}$, respectively.
Among the 46 cores, we found four massive cores associated with YSOs.
The cores have a large mass of $\gtsim1000M_{\odot}$, a line width of $\gtsim 2.5$ km s$^{-1}$, and an excitation temperature of $\gtsim10$ K.

\item
We investigated the fractional abundance of N$_2$H$^{+}$ for all of the cores.
The values are estimated to be in the range $(4-10) \times 10^{-10}$, indicating that there is a large variation among the cores.

\item
We estimated  the average star formation efficiency of M17 SWex to be $\sim 9-17\%$.

\item
Comparison of the virial mass and the core mass estimated from the Herschel data revealed 
that most of the cores can be dynamically stable if we assume a large external pressure.
For some cores, the results of the comparison indicate that internal pressure due to turbulence is small,
and they can collapse by the self-gravity if there is no internal supporting force by the magnetic field.

\item
Assuming a radial density profile of spherical gas, we examined the $N$(H$_2$) distribution 
to infer the index $\alpha$ of the density profile of the cores $\rho (r) \propto r^{-\alpha}$.
We found that 32 out of 46 cores have a value of $\alpha$ in the range $1.3-2.2$,
suggesting that the cores are gravitationally stable or free-falling.

\item
We estimated that the magnetic field of $\sim 100-600$ $\mu$G is required for the cores to be critical.
This value is comparable to the strength of the magnetic filed recently measured by \cite{Sugitani2019}.
Previous studies \cite[e.g.,][]{Povich2010} have shown that there is no active massive star formation in M17 SWex, 
and we suggest that it is due to the cloud-supporting force by the magnetic field
which prevents the cores from collapsing.
\end{enumerate}

%%%%%%%%%%%%%%%%%%%%
%             Acknowledgements              %
%%%%%%%%%%%%%%%%%%%%

\begin{ack}
We thank Dr. M. S. Povich for his reviewing this paper and for a number of useful comments.
We are also grateful to the other members of Star Formation Legacy project for their support
during the observations.
This work was supported by JSPS KAKENHI Grant Numbers JP17K00963, JP17H02863, JP17H01118, JP16K12749,
and NAOJ ALMA Scientific Research Grant Numbers 2017-04A. 
YS also received support from the ANR (project NIKA2SKY, grant agreement ANR-15-CE31-0017).
The 45 m radio telescope is operated by NRO, a branch of NAOJ. 
\end{ack}

%%%%%%%%%%%%%%%%%%%%%
%                            References             %
%%%%%%%%%%%%%%%%%%%%%

%\input{references}
%\bibliography{M17}

%%%%%%%%%%%%%%%%%%%%%
%                           Tables                       %
%%%%%%%%%%%%%%%%%%%%%
%\input{table}
%\documentclass[]{pasj01}
%\usepackage{lscape}

%\begin{document}

% Table 1%%%%%%%%%%%%%%%%%%%%%%%%%%%%%%%%%%%%%%%%%%%%%%%%%%%%%%%
\clearpage

\begin{table}
\caption{The observed molecular lines} 
\begin{tabular}{lcrcc}  
\hline\noalign{\vskip3pt} 
	\multicolumn{1}{c}{Molecule} & \multicolumn{1}{c}{Transition} & \multicolumn{1}{c}{Frequency} & \multicolumn{1}{c}{Beam} & \multicolumn{1}{c}{$\Delta T_{\rm mb}$}\\
	\multicolumn{1}{c}{} & \multicolumn{1}{c}{} & \multicolumn{1}{c}{(GHz)} & \multicolumn{1}{c}{} & \multicolumn{1}{c}{(K)} \\  [2pt] 
\hline\noalign{\vskip3pt} 
N$_2$H$^{+}$	&	$J=1-0$				&	  93.1737637 	&	$17\farcs9$	&	0.44	\\
CCS			&	$J_{N}$=8$_{7}-7_{6}$	&	  93.8700980 	&	$17\farcs8$	&	0.42	\\
C$^{18}$O	&	$J=1-0$				&	109.7821760 	&	$15\farcs2$	&	0.43	\\
$^{13}$CO	&	$J=1-0$				&	110.2013540 	&	$15\farcs1$	&	0.37	\\
$^{12}$CO	&	$J=1-0$				&	115.2712040 	&	$14\farcs5$	&	0.90	\\
\hline
\end{tabular} \label{tab:line}
   \begin{tabnote}
The rest frequency for the N$_2$H$^{+}$ line is taken from \cite{Pagani2009}, and those of the other lines are taken from the website of \cite{Lovas}.
   \end{tabnote}
\end{table} 
\clearpage

% Table 2%%%%%%%%%%%%%%%%%%%%%%%%%%%%%%%%%%%%%%%%%%%%%%%%%%%%%%%
%\begin{landscape}
\renewcommand{\arraystretch}{0.7}
\begin{table}
\caption{The cores and associated YSOs} 
\setlength{\tabcolsep}{0.04in} 
\begin{tabular}{cccccc} 
\hline
\multicolumn{3}{c}{}&\multicolumn{3}{c}{number of YSOs\footnotemark[(a)]}\\
\cline{4-6}
Core	&	$l$ &	$b$ &	Stage 0/I	&	Stage II/III	&	A\footnotemark[$*$]	\\
	&({$^\circ$})	&	({$^\circ$}) 	&&&	\\	
	\hline
1	&	13.84 	&	-0.49 	&	1	&	7	&	0	\\
2	&	13.88 	&	-0.59 	&	2	&	1	&	0	\\
3	&	13.90 	&	-0.51 	&	10	&	2	&	0	\\
4	&	13.91 	&	-0.45 	&	2	&	0	&	0	\\
5	&	13.94 	&	-0.44 	&	0	&	2	&	0	\\
6	&	13.95 	&	-0.41 	&	0	&	3	&	2	\\
7	&	13.96 	&	-0.44 	&	0	&	2	&	2	\\
8	&	13.97 	&	-0.46 	&	4	&	0	&	1	\\
9	&	13.97 	&	-0.41 	&	0	&	0	&	1	\\
10	&	13.97 	&	-0.39 	&	3	&	1	&	0	\\
11	&	14.00 	&	-0.56 	&	3	&	2	&	0	\\
12	&	14.02 	&	-0.51 	&	4	&	2	&	2	\\
13	&	14.09 	&	-0.56 	&	1	&	0	&	0	\\
14	&	14.09 	&	-0.52 	&	0	&	0	&	0	\\
15	&	14.11 	&	-0.57 	&	9	&	3	&	4	\\
16	&	14.13 	&	-0.52 	&	4	&	6	&	0	\\
17	&	14.15 	&	-0.55 	&	1	&	1	&	2	\\
18	&	14.15 	&	-0.57 	&	1	&	0	&	0	\\
19	&	14.17 	&	-0.61 	&	3	&	0	&	1	\\
20	&	14.18 	&	-0.53 	&	5	&	4	&	1	\\
21	&	14.19 	&	-0.59 	&	0	&	2	&	1	\\
22	&	14.19 	&	-0.51 	&	2	&	3	&	1	\\
23	&	14.23 	&	-0.51 	&	11	&	8	&	1	\\
24	&	14.23 	&	-0.57 	&	0	&	1	&	1	\\
25	&	14.23 	&	-0.60 	&	8	&	6	&	4	\\
26	&	14.27 	&	-0.56 	&	1	&	0	&	0	\\
27	&	14.28 	&	-0.58 	&	3	&	1	&	0	\\
28	&	14.29 	&	-0.66 	&	0	&	1	&	0	\\
29	&	14.29 	&	-0.51 	&	2	&	1	&	3	\\
30	&	14.30 	&	-0.59 	&	1	&	2	&	0	\\
31	&	14.31 	&	-0.49 	&	1	&	0	&	0	\\
32	&	14.33 	&	-0.53 	&	0	&	0	&	0	\\
33	&	14.33 	&	-0.65 	&	14	&	11	&	4	\\
34	&	14.36 	&	-0.72 	&	0	&	1	&	0	\\
35	&	14.37 	&	-0.51 	&	0	&	0	&	0	\\
36	&	14.37 	&	-0.73 	&	1	&	1	&	0	\\
37	&	14.44 	&	-0.70 	&	1	&	2	&	0	\\
38	&	14.45 	&	-0.76 	&	1	&	1	&	0	\\
39	&	14.45 	&	-0.61 	&	1	&	1	&	0	\\
40	&	14.47 	&	-0.61 	&	1	&	0	&	0	\\
41	&	14.51 	&	-0.60 	&	0	&	0	&	0	\\
42	&	14.55 	&	-0.62 	&	0	&	0	&	0	\\
43	&	14.60 	&	-0.62 	&	0	&	2	&	0	\\
44	&	14.60 	&	-0.55 	&	1	&	1	&	0	\\
45	&	14.61 	&	-0.60 	&	0	&	0	&	0	\\
46	&	14.64 	&	-0.57 	&	6	&	7	&	0	\\
\hline

\end{tabular}
 \label{tab:core}
   \begin{tabnote}
{\footnotemark[$(a)$]}\cite{Povich2016}
{\footnotemark[$*$]} Ambiguous stage
   \end{tabnote}
\end{table}
%\end{landscape}
\clearpage

% Table 3%%%%%%%%%%%%%%%%%%%%%%%%%%%%%%%%%%%%%%%%%%%%%%%%%%%%%%%

\begin{table}
\caption{The line parameters of N${_2}$H$^{+}$} 
\begin{tabular}{cccccc}  
\hline\noalign{\vskip3pt} 

Core	&	$V_{\rm {LSR}}$			&	{$\tau_{\rm {tot}}$} 			&	${\Delta}V$			&	{$T^{\rm{N{_2}H^{+}}}_{\rm ex}$} 			&	$N$(N${_2}$H$^{+}$)	\\
	&	{(km$^{-1}$)} 			&				&	{(km$^{-1}$)} 			&	(K)			&	{(cm$^{-2}$)} 	\\
\hline

1	&	24.40 	$\pm$	0.04 	&	5.7 	$\pm$	3.0 	&	0.85 	$\pm$	0.10 	&	4.39 	$\pm$	0.44 	&	1.29E+13	\\
2	&	18.42 	$\pm$	0.04 	&	6.2 	$\pm$	0.8 	&	1.07 	$\pm$	0.10 	&	4.45 			&	1.81E+13	\\
3	&	22.86 	$\pm$	0.03 	&	4.7 	$\pm$	0.9 	&	1.74 	$\pm$	0.09 	&	5.72 	$\pm$	0.26 	&	3.23E+13	\\
4	&	22.77 	$\pm$	0.05 	&	1.6 	$\pm$	1.3 	&	1.83 	$\pm$	0.20 	&	5.92 	$\pm$	1.79 	&	1.25E+13	\\
5	&	22.30 	$\pm$	0.11 	&	5.4 	$\pm$	3.3 	&	2.11 	$\pm$	0.37 	&	3.58 	$\pm$	0.20 	&	2.27E+13	\\
6	&	19.87 	$\pm$	0.07 	&	5.8 	$\pm$	0.8 	&	2.41 	$\pm$	0.19 	&	3.93 			&	3.16E+13	\\
7	&	22.97 	$\pm$	0.04 	&	13.2 	$\pm$	6.6 	&	0.82 	$\pm$	0.13 	&	3.86 	$\pm$	0.15 	&	2.39E+13	\\
8	&	21.92 	$\pm$	0.04 	&	9.2 	$\pm$	4.8 	&	0.70 	$\pm$	0.09 	&	3.96 	$\pm$	0.26 	&	1.48E+13	\\
9	&	19.64 	$\pm$	0.14 	&	8.1 	$\pm$	3.9 	&	2.66 	$\pm$	0.38 	&	3.31 	$\pm$	0.08 	&	3.90E+13	\\
10	&	20.88 	$\pm$	0.20 	&	1.4 	$\pm$	1.8 	&	4.00 	$\pm$	0.60 	&	4.00 	$\pm$	1.19 	&	1.36E+13	\\
11	&	18.18 	$\pm$	0.04 	&	4.7 	$\pm$	1.1 	&	1.87 	$\pm$	0.12 	&	4.52 	$\pm$	0.19 	&	2.43E+13	\\
12	&	20.06 	$\pm$	0.02 	&	3.6 	$\pm$	0.9 	&	1.32 	$\pm$	0.08 	&	6.17 	$\pm$	0.50 	&	2.09E+13	\\
13	&	22.08 	$\pm$	0.03 	&	1.7 	$\pm$	0.7 	&	2.15 	$\pm$	0.13 	&	6.32 	$\pm$	1.02 	&	1.72E+13	\\
14	&	22.10 	$\pm$	0.03 	&	5.2 	$\pm$	2.5 	&	0.90 	$\pm$	0.10 	&	4.53 	$\pm$	0.45 	&	1.29E+13	\\
15	&	19.90 	$\pm$	0.02 	&	1.9 	$\pm$	0.2 	&	3.19 	$\pm$	0.05 	&	10.49 	$\pm$	0.58 	&	6.36E+13	\\
16	&	21.22 	$\pm$	0.02 	&	6.0 	$\pm$	1.1 	&	1.10 	$\pm$	0.06 	&	5.51 	$\pm$	0.24 	&	2.43E+13	\\
17	&	21.30 	$\pm$	0.03 	&	2.2 	$\pm$	1.0 	&	1.38 	$\pm$	0.09 	&	6.39 	$\pm$	1.05 	&	1.44E+13	\\
18	&	21.24 	$\pm$	0.03 	&	3.7 	$\pm$	1.3 	&	1.40 	$\pm$	0.12 	&	4.66 	$\pm$	0.37 	&	1.51E+13	\\
19	&	20.14 	$\pm$	0.03 	&	8.3 	$\pm$	5.5 	&	0.59 	$\pm$	0.09 	&	3.94 	$\pm$	0.37 	&	1.11E+13	\\
20	&	21.46 	$\pm$	0.02 	&	1.5 	$\pm$	0.4 	&	1.92 	$\pm$	0.06 	&	9.83 	$\pm$	1.29 	&	2.67E+13	\\
21	&	19.02 	$\pm$	0.07 	&	3.6 	$\pm$	1.7 	&	2.17 	$\pm$	0.25 	&	3.93 	$\pm$	0.29 	&	1.78E+13	\\
22	&	20.36 	$\pm$	0.04 	&	2.8 	$\pm$	1.0 	&	1.82 	$\pm$	0.14 	&	5.30 	$\pm$	0.54 	&	1.80E+13	\\
23	&	19.55 	$\pm$	0.01 	&	2.3 	$\pm$	0.2 	&	2.66 	$\pm$	0.05 	&	11.34 	$\pm$	0.56 	&	7.51E+13	\\
24	&	21.80 	$\pm$	0.08 	&	5.7 	$\pm$	3.5 	&	1.51 	$\pm$	0.28 	&	3.63 	$\pm$	0.22 	&	1.77E+13	\\
25	&	19.87 	$\pm$	0.06 	&	1.6 	$\pm$	1.0 	&	2.62 	$\pm$	0.23 	&	4.97 	$\pm$	0.96 	&	1.37E+13	\\
26	&	19.41 	$\pm$	0.04 	&	1.5 	$\pm$	0.1 	&	2.11 	$\pm$	0.14 	&	6.08 			&	1.36E+13	\\
27	&	20.85 	$\pm$	0.04 	&	2.3 	$\pm$	1.0 	&	1.66 	$\pm$	0.13 	&	6.06 	$\pm$	0.96 	&	1.64E+13	\\
28	&	19.10 	$\pm$	0.04 	&	12.6 	$\pm$	10.1 	&	0.53 	$\pm$	0.11 	&	3.46 	$\pm$	0.20 	&	1.28E+13	\\
29	&	19.77 	$\pm$	0.03 	&	5.6 	$\pm$	2.0 	&	0.90 	$\pm$	0.08 	&	4.81 	$\pm$	0.36 	&	1.54E+13	\\
30	&	21.28 	$\pm$	0.04 	&	9.5 	$\pm$	4.5 	&	0.75 	$\pm$	0.10 	&	3.74 	$\pm$	0.18 	&	1.51E+13	\\
31	&	18.33 	$\pm$	0.02 	&	3.4 	$\pm$	2.5 	&	0.62 	$\pm$	0.06 	&	5.53 	$\pm$	1.38 	&	7.94E+12	\\
32	&	19.93 	$\pm$	0.02 	&	8.4 	$\pm$	1.8 	&	1.03 	$\pm$	0.07 	&	4.94 	$\pm$	0.17 	&	2.75E+13	\\
33	&	22.32 	$\pm$	0.01 	&	1.0 	$\pm$	0.1 	&	3.48 	$\pm$	0.03 	&	31.68 	$\pm$	2.50 	&	2.71E+14	\\
34	&	20.46 	$\pm$	0.04 	&	1.4 	$\pm$	2.2 	&	0.95 	$\pm$	0.12 	&	6.59 	$\pm$	4.89 	&	6.59E+12	\\
35	&	15.30 	$\pm$	0.96 	&	16.2 	$\pm$	49.0 	&	0.96 	$\pm$	0.35 	&	3.00	 	&	9.59E+13	\\
36	&	20.08 	$\pm$	0.19 	&	0.3 	$\pm$	0.0 	&	3.45 	$\pm$	0.52 	&	6.95 			&	5.16E+12	\\
37	&	17.58 	$\pm$	0.04 	&	3.2 	$\pm$	1.3 	&	1.54 	$\pm$	0.14 	&	5.08 	$\pm$	0.54 	&	1.63E+13	\\
38	&	21.35 	$\pm$	0.06 	&	3.1 	$\pm$	2.2 	&	1.37 	$\pm$	0.20 	&	4.60 	$\pm$	0.80 	&	1.19E+13	\\
39	&	20.35 	$\pm$	0.02 	&	9.6 	$\pm$	2.6 	&	0.76 	$\pm$	0.06 	&	4.80 	$\pm$	0.21 	&	2.21E+13	\\
40	&	19.07 	$\pm$	0.08 	&	1.6 	$\pm$	0.3 	&	1.61 	$\pm$	0.27 	&	4.80 			&	7.70E+12	\\
41	&	17.51 	$\pm$	0.12 	&	1.2 	$\pm$	0.1 	&	2.70 	$\pm$	0.36 	&	4.56 			&	8.69E+12	\\
42	&	18.62 	$\pm$	0.05 	&	3.5 	$\pm$	0.5 	&	1.02 	$\pm$	0.14 	&	4.56 			&	1.00E+13	\\
43	&	19.15 	$\pm$	0.07 	&	2.6 	$\pm$	2.2 	&	1.51 	$\pm$	0.24 	&	4.55 	$\pm$	0.95 	&	1.10E+13	\\
44	&	17.66 	$\pm$	0.04 	&	1.0 	$\pm$	0.8 	&	2.13 	$\pm$	0.15 	&	9.03 	$\pm$	3.79 	&	1.77E+13	\\
45	&	19.32 	$\pm$	0.05 	&	2.8 	$\pm$	0.4 	&	1.28 	$\pm$	0.17 	&	4.56 			&	1.00E+13	\\
46	&	18.74 	$\pm$	0.01 	&	2.0 	$\pm$	0.2 	&	2.90 	$\pm$	0.05 	&	13.04 	$\pm$	0.74 	&	9.10E+13	\\

\hline
\end{tabular} \label{tab:line_n2hp}
   \begin{tabnote}
The parameters are measured toward the peak position of the cores (table \ref{tab:core}).
$T^{\rm{N{_2}H^{+}}}_{\rm ex}$ values without an error are the values derive by interpolating the $T^{\rm{N{_2}H^{+}}}_{\rm ex}$ values of the neighboring pixels.
   \end{tabnote}
\end{table} 

\clearpage
% Table 4%%%%%%%%%%%%%%%%%%%%%%%%%%%%%%%%%%%%%%%%%%%%%%%%%%%%%%%

\begin{table}
\caption{The physical parameters of the cores} 
\begin{tabular}{cccrrr}  
\hline\noalign{\vskip3pt} 

Core	&	$S$	&	$R_{\rm core}$	&	$M_{\rm{core}}$			&	$M_{\rm{vir}}$	& {$\Sigma M_\star$} \\
	&	{(arcmin$^{2}$)} 	&	{(pc)}	&	{($M_{\odot}$)}		&	{($M_{\odot}$)}	& {($M_{\odot}$)} \\
\hline

1	&	1.44 	&	0.40 	&	272 	$\pm$	2 	&	59 	&	103 	\\
2	&	0.72 	&	0.28 	&	83 	$\pm$	2 	&	67 	&	39 	\\
3	&	3.94 	&	0.65 	&	717 	$\pm$	4 	&	412 	&	154 	\\
4	&	2.02 	&	0.47 	&	275 	$\pm$	3 	&	328 	&	26 	\\
5	&	0.77 	&	0.29 	&	90 	$\pm$	2 	&	268 	&	26 	\\
6	&	1.69 	&	0.43 	&	269 	$\pm$	2 	&	522 	&	64 	\\
7	&	0.50 	&	0.23 	&	85 	$\pm$	1 	&	33 	&	51 	\\
8	&	1.39 	&	0.39 	&	155 	$\pm$	2 	&	40 	&	64 	\\
9	&	0.44 	&	0.22 	&	62 	$\pm$	1 	&	323 	&	13 	\\
10	&	1.06 	&	0.34 	&	110 	$\pm$	2 	&	1139 	&	51 	\\
11	&	1.11 	&	0.35 	&	156 	$\pm$	2 	&	255 	&	64 	\\
12	&	1.22 	&	0.36 	&	267 	$\pm$	2 	&	132 	&	103 	\\
13	&	0.70 	&	0.28 	&	209 	$\pm$	2 	&	267 	&	13 	\\
14	&	0.50 	&	0.23 	&	57 	$\pm$	1 	&	39 	&	0 	\\
15	&	2.27 	&	0.50 	&	1007 	$\pm$	3 	&	1054 	&	206 	\\
16	&	2.05 	&	0.47 	&	416 	$\pm$	3 	&	119 	&	129 	\\
17	&	1.17 	&	0.36 	&	266 	$\pm$	2 	&	141 	&	51 	\\
18	&	0.48 	&	0.23 	&	125 	$\pm$	1 	&	95 	&	13 	\\
19	&	0.42 	&	0.21 	&	54 	$\pm$	1 	&	16 	&	51 	\\
20	&	1.30 	&	0.38 	&	387 	$\pm$	2 	&	289 	&	129 	\\
21	&	1.11 	&	0.35 	&	158 	$\pm$	2 	&	343 	&	39 	\\
22	&	0.77 	&	0.29 	&	274 	$\pm$	2 	&	201 	&	77 	\\
23	&	2.52 	&	0.52 	&	1152 	$\pm$	3 	&	774 	&	257 	\\
24	&	0.98 	&	0.33 	&	108 	$\pm$	2 	&	157 	&	26 	\\
25	&	3.36 	&	0.60 	&	541 	$\pm$	3 	&	869 	&	232 	\\
26	&	0.70 	&	0.28 	&	106 	$\pm$	2 	&	258 	&	13 	\\
27	&	0.45 	&	0.22 	&	79 	$\pm$	1 	&	128 	&	51 	\\
28	&	0.39 	&	0.21 	&	74 	$\pm$	1 	&	12 	&	13 	\\
29	&	2.70 	&	0.54 	&	405 	$\pm$	3 	&	92 	&	77 	\\
30	&	0.64 	&	0.26 	&	71 	$\pm$	1 	&	31 	&	39 	\\
31	&	0.48 	&	0.23 	&	74 	$\pm$	1 	&	19 	&	13 	\\
32	&	1.06 	&	0.34 	&	179 	$\pm$	2 	&	76 	&	0 	\\
33	&	9.80 	&	1.03 	&	3026 	$\pm$	6 	&	2617 	&	373 	\\
34	&	0.70 	&	0.28 	&	122 	$\pm$	2 	&	53 	&	13 	\\
35	&	0.39 	&	0.21 	&	29 	$\pm$	1 	&	1075 	&	0 	\\
36	&	0.78 	&	0.29 	&	140 	$\pm$	2 	&	727 	&	26 	\\
37	&	1.48 	&	0.40 	&	300 	$\pm$	2 	&	199 	&	39 	\\
38	&	0.66 	&	0.27 	&	120 	$\pm$	1 	&	105 	&	26 	\\
39	&	0.56 	&	0.25 	&	107 	$\pm$	1 	&	30 	&	26 	\\
40	&	0.48 	&	0.23 	&	76 	$\pm$	1 	&	124 	&	13 	\\
41	&	0.39 	&	0.21 	&	47 	$\pm$	1 	&	314 	&	0 	\\
42	&	1.52 	&	0.41 	&	239 	$\pm$	2 	&	89 	&	0 	\\
43	&	0.39 	&	0.21 	&	43 	$\pm$	1 	&	98 	&	26 	\\
44	&	0.89 	&	0.31 	&	218 	$\pm$	2 	&	294 	&	26 	\\
45	&	0.41 	&	0.21 	&	57 	$\pm$	1 	&	72 	&	0 	\\
46	&	4.38 	&	0.69 	&	1488 	$\pm$	4 	&	1216 	&	167 	\\

\hline
\end{tabular} \label{tab:mass}
  \begin{tabnote}
The parameters $S$, $R_{\rm core}$, $M_{\rm{core}}$, and $M_{\rm{vir}}$ are derived
on the assumption of a uniform distance of 2.0 kpc.
   \end{tabnote}

\end{table} 

\clearpage

% Table 4%%%%%%%%%%%%%%%%%%%%%%%%%%%%%%%%%%%%%%%%%%%%%%%%%%%%%%%

\begin{table}
\caption{The fractional abundance of the cores} 
\begin{tabular}{cccc}  
\hline
Core	&	$s$			&	$i$			&	$\gamma$	\\
	&	   			&	(cm$^{-2}$)	&		\\
\hline
1	&	1.7E-10	$\pm$	0.8E-10	&	7.66E+12	$\pm$	2.20E+12	&	0.23 	\\
2	&	6.0E-10	$\pm$	2.0E-10	&	4.43E+12	$\pm$	3.06E+12	&	0.42 	\\
3	&	4.1E-10	$\pm$	0.3E-10	&	6.85E+12	$\pm$	0.83E+12	&	0.65 	\\
4	&	3.6E-10	$\pm$	0.4E-10	&	4.30E+12	$\pm$	0.82E+12	&	0.63 	\\
5	&	1.7E-09	$\pm$	0.9E-09	&	-9.49E+12	$\pm$	1.39E+13	&	0.30 	\\
6	&	1.0E-10	$\pm$	0.6E-10	&	9.80E+12	$\pm$	1.28E+12	&	0.20 	\\
7	&	3.0E-10	$\pm$	1.7E-10	&	6.49E+12	$\pm$	3.80E+12	&	0.32 	\\
8	&	1.8E-10	$\pm$	2.4E-10	&	1.08E+13	$\pm$	0.35E+13	&	0.10 	\\
9	&	1.8E-09	$\pm$	0.7E-09	&	-1.42E+13	$\pm$	1.39E+13	&	0.52 	\\
10	&	9.9E-10	$\pm$	2.8E-10	&	5.65E+12	$\pm$	4.00E+12	&	0.43 	\\
11	&	1.9E-10	$\pm$	0.9E-10	&	1.15E+13	$\pm$	0.18E+13	&	0.27 	\\
12	&	3.2E-10	$\pm$	0.9E-10	&	4.91E+12	$\pm$	2.82E+12	&	0.40 	\\
13	&	-3.1E-10	$\pm$	7.9E-11	&	2.95E+13	$\pm$	0.31E+13	&	-0.52 	\\
14	&	9.1E-10	$\pm$	2.0E-10	&	-2.75E+12	$\pm$	3.08E+12	&	0.63 	\\
15	&	3.6E-10	$\pm$	0.1E-10	&	7.78E+12	$\pm$	0.87E+12	&	0.92 	\\
16	&	4.2E-10	$\pm$	0.5E-10	&	5.74E+12	$\pm$	1.26E+12	&	0.63 	\\
17	&	4.8E-10	$\pm$	0.5E-10	&	4.75E+12	$\pm$	1.38E+12	&	0.63 	\\
18	&	2.7E-10	$\pm$	5.2E-10	&	0.89E+13	$\pm$	1.75E+13	&	0.10 	\\
19	&	0.8E-10	$\pm$	2.3E-10	&	9.48E+12	$\pm$	4.12E+12	&	0.08 	\\
20	&	2.9E-10	$\pm$	0.3E-10	&	4.87E+12	$\pm$	1.10E+12	&	0.79 	\\
21	&	2.2E-10	$\pm$	1.0E-10	&	5.56E+12	$\pm$	1.94E+12	&	0.27 	\\
22	&	3.9E-10	$\pm$	0.3E-10	&	-1.75E+12	$\pm$	1.27E+12	&	0.91 	\\
23	&	3.4E-10	$\pm$	0.1E-10	&	5.06E+12	$\pm$	0.89E+12	&	0.91 	\\
24	&	-1.1E-09	$\pm$	2.7E-10	&	2.72E+13	$\pm$	0.34E+13	&	-0.46 	\\
25	&	5.6E-11	$\pm$	3.9E-11	&	9.45E+12	$\pm$	0.70E+12	&	0.11 	\\
26	&	4.1E-10	$\pm$	2.9E-10	&	1.00E+13	$\pm$	0.58E+13	&	0.21 	\\
27	&	1.3E-10	$\pm$	1.2E-10	&	1.25E+13	$\pm$	0.28E+13	&	0.19 	\\
28	&	-2.0E-09	$\pm$	3.6E-09	&	7.16E+13	$\pm$	9.02E+13	&	-0.12 	\\
29	&	1.8E-10	$\pm$	0.5E-10	&	6.38E+12	$\pm$	1.06E+12	&	0.33 	\\
30	&	5.0E-10	$\pm$	1.9E-10	&	1.99E+12	$\pm$	2.87E+12	&	0.40 	\\
31	&	6.8E-10	$\pm$	2.3E-10	&	-1.04E+12	$\pm$	4.72E+12	&	0.42 	\\
32	&	4.8E-10	$\pm$	1.0E-10	&	5.83E+12	$\pm$	2.33E+12	&	0.54 	\\
33	&	9.1E-10	$\pm$	0.1E-10	&	-6.41E+12	$\pm$	5.79E+11	&	0.96 	\\
34	&	4.9E-10	$\pm$	1.7E-10	&	1.27E+12	$\pm$	4.04E+12	&	0.40 	\\
35	&	9.4E-09	$\pm$	3.8E-09	&	-8.23E+13	$\pm$	3.55E+13	&	0.62 	\\
36	&	-1.2E-09	$\pm$	3.7E-10	&	4.27E+13	$\pm$	0.87E+13	&	-0.45 	\\
37	&	8.8E-11	$\pm$	8.8E-11	&	1.06E+13	$\pm$	0.24E+13	&	0.11 	\\
38	&	-3.3E-10	$\pm$	1.2E-10	&	2.44E+13	$\pm$	0.31E+13	&	-0.41 	\\
39	&	1.3E-11	$\pm$	2.1E-10	&	1.68E+13	$\pm$	0.53E+13	&	0.01 	\\
40	&	1.8E-11	$\pm$	2.3E-10	&	1.22E+13	$\pm$	0.47E+13	&	0.01 	\\
41	&	1.5E-09	$\pm$	5.6E-10	&	-1.15E+13	$\pm$	0.89E+13	&	0.53 	\\
42	&	-2.5E-08	$\pm$	6.1E-09	&	5.83E+14	$\pm$	1.29E+14	&	-0.41 	\\
43	&	7.0E-10	$\pm$	1.8E-10	&	-1.02E+12	$\pm$	2.59E+12	&	0.65 	\\
44	&	6.8E-10	$\pm$	0.8E-10	&	-3.84E+12	$\pm$	2.72E+12	&	0.74 	\\
45	&	5.7E-10	$\pm$	0.9E-10	&	-9.70E+11	$\pm$	1.68E+12	&	0.79 	\\
46	&	6.0E-10	$\pm$	0.2E-10	&	3.54E+12	$\pm$	1.07E+12	&	0.86 	\\
\hline
\end{tabular} \label{tab:fractional}
\end{table} 

%\end{document}

%%%%%%%%%%%%%%%%%%%%%
%                           figures                      %
%%%%%%%%%%%%%%%%%%%%%
%\input{figure}
%\documentclass{aastex}
%\begin{document}

%%Fig 1%%%%%%%%%%%%%%%%%%%%%%%%%%%%%%%%%%

%%Fig 2%%%%%%%%%%%%%%%%%%%%%%%%%%%%%%%%%%

%%Fig 3%%%%%%%%%%%%%%%%%%%%%%%%%%%%%%%%%%

%%Fig 4%%%%%%%%%%%%%%%%%%%%%%%%%%%%%%%%%%

%%Fig 5%%%%%%%%%%%%%%%%%%%%%%%%%%%%%%%%%%

%%Fig 6%%%%%%%%%%%%%%%%%%%%%%%%%%%%%%%%%%

%%Fig 7%%%%%%%%%%%%%%%%%%%%%%%%%%%%%%%%%%

%%Fig 8%%%%%%%%%%%%%%%%%%%%%%%%%%%%%%%%%%

%%Fig 9%%%%%%%%%%%%%%%%%%%%%%%%%%%%%%%%%%

%%Fig 10%%%%%%%%%%%%%%%%%%%%%%%%%%%%%%%%%%

%%Fig 12%%%%%%%%%%%%%%%%%%%%%%%%%%%%%%%%%%

%%Fig 14%%%%%%%%%%%%%%%%%%%%%%%%%%%%%%%%%%

%\end{document}

\end{document}